# *PT*-symmetric chiral metamaterials: Asymmetric effects and *PT*-phase control

Ioannis Katsantonis[1,2*], Sotiris Droulias[1,2†], Costas M. Soukoulis[1,3], Eleftherios N. Economou[1,4] and Maria Kafesaki[1,2‡]
[1]*Institute of Electronic Structure and Laser, Foundation for Research and Technology Hellas, 70013, Heraklion, Crete, Greece*
[2]*University of Crete, Department of Materials Science and Technology, 70013, Heraklion, Greece*
[3]*Ames Laboratory and Department of Physics and Astronomy, Iowa States, Ames. Iowa 50010, USA*
[4]*University of Crete, Department of Physics, 70013, Heraklion, Greece*

We investigate the influence of chirality on the *PT*-symmetric and *PT*-broken phase of *PT*-symmetric chiral systems. Starting from the point that transverse magnetic (TM) and transverse electric (TE) waves have different exceptional points, we show that with circularly polarized waves (which are linear combinations of TM and TE waves) mixed *PT*-symmetric phases can be realized and the extent of these phases can be highly controlled by either or both the chirality and the angle of incidence. Additionally, while the transmission of both TM and TE waves in non-chiral *PT*-symmetric systems is the same for forward and backward propagation, we show that with chirality this symmetry can be broken. As a result, it is possible to realize asymmetric, i.e. side-dependent, rotation and ellipticity in the polarization state of the transmitted wave. Our results constitute a simple example of a chiral *PT*-symmetric optical system in which the various phases (full *PT*, mixed, broken) and the asymmetric effects can be easily tuned by adjusting the chirality parameter and/or the angle of incidence.

## I. INTRODUCTION

One of the most fundamental axioms in Quantum mechanics is that all physical observables should correspond to real eigenvalues, so the use of Hermitian Hamiltonians guarantees that the entire eigenspectrum of the system will be real. The hermiticity of the Hamiltonian though is not necessary for the reality of the eigenvalues. Parity-Time (*PT*-) symmetric Hamiltonians are a class of Hamiltonians which, despite of being non-Hermitian, present the possibility of real eigenvalues [1-4], as was first shown in 1998, by Bender and co-workers [1].

In principle, $PT$ -symmetric Hamiltonian means that it is invariant under the combined action of the parity, $\hat{P}$, and time-reversal, $\hat{T}$, operators, i.e. $[\hat{P}\hat{T},\hat{H}]=0$, with the $[\ldots]$ denoting the commutator. Since the action of parity on the momentum and position operators, $\hat{p}$ and $\hat{r}$ respectively, results to $\hat{p}\rightarrow-\hat{p},\ \hat{r}\rightarrow-\hat{r}$ , and the time-reversal to $\hat{p}\rightarrow-\hat{p},\ \hat{r}\rightarrow\hat{r}$ and $i\rightarrow-i$ [4], the requirement of *PT*-symmetry in a single particle, one-dimensional quantum Hamiltonian results to the symmetry condition for the potential $V^*(-r)=V(r)$. The reality of the eigenvalues is obtained for sure if both the Hamiltonian and its eigenstates obey the *PT*-symmetry. In fact a characteristic feature in *PT*-symmetric Hamiltonians is the reality of the eigenvalues below a critical value of the potential (region where also eigenstates are *PT*-symmetric – usually called *PT*-phase), while above that critical value (in the so-called *PT*-broken phase) the *PT* symmetry of the eigenstates breaks down and the eigenvalues become complex. At the transition point between *PT* and *PT*-broken phase, the so-called exceptional point (EP), two or more eigenvalues and eigenvectors coincide, making EP to be a singular point, associated with peculiar dynamic evolution characteristics.

Recently, the *PT*-symmetry concept was extensively applied in optical physics [5-12]. Realizing *PT*-symmetry in optical systems requires $n^*(-r)=n(r)$, for the refractive index $n$ (in the framework of paraxial approximation of wave equation) and it is attained in practice by spatially modulating gain and loss in materials. Since gain and loss can be to a large extent controlled externally, the application of *PT*-symmetry in optics offered the opportunity to demonstrate also experimentally many unique and novel phenomena associated with *PT*-symmetric systems. Such phenomena include anisotropic transmission resonances, i.e. unidirectional reflectionless perfect transmission [13], unidirectional invisibility [14], *PT*-breaking transitions [15], coherent perfect absorption (CPA) [16] and extraordinary nonlinear effects [17]. More recently, a lot of work appeared combining *PT*-symmetry with metamaterials, and mainly with zero index metamaterials [18-22], showing particularly interesting effects; for example tunnelling [19] mediated by excitation of a surface wave at the interface between the gain and loss domains, unidirectional transparency, asymmetric reflection and other asymmetric propagation features, coherent perfect absorber-laser modes [23], etc.

Many of the observed and demonstrated interesting phenomena in *PT*-symmetric systems (such as coherent perfect absorption or lasing) are observed at or beyond the exceptional points (EPs) at which the reality of the spectrum breaks down and the *PT*-symmetry of the eigensolutions vanishes, although the "Hamiltonian" continues to be *PT*-symmetric. This in optical systems (where the equivalent "potential" involves the frequency and the refractive index) occurs at a critical value of either frequency or refractive index. A lot of work up to now has been devoted to the identification of exceptional points in different systems and the investigation of novel effects related with them [24]. A large amount of such work concerns scattering configurations rather than paraxial beam propagation systems. There, it was shown that the identification of EPs and the different *PT*-related phases can be done through the eigenvalues of the scattering matrix, which in the *PT*-phase are unimodular while in the *PT*-broken phase are not unimodular [13]. Examining the scattering matrix eigenvalues and eigenvectors in *PT*-symmetric structures under oblique incidence, an interesting feature was observed, namely different phase transition points for transverse electric (TE) polarization and for transverse magnetic (TM) polarization under the same angle of incidence [25]. This resulted to a mixed *PT*-phase for circularly polarized waves (formed by a linear combination of TM and TE waves) and opened the possibility for multiple exceptional points even in simple two-dimensional systems, expanding thus the potential of *PT*-symmetric systems for novel and unusual propagation and scattering features.

As was mentioned already, a lot of work on $PT$-symmetry has combined the $PT$-symmetry concept with metamaterials, investigating the unique effects that can arise by combining the uncommon and fascinating properties of metamaterials (e.g. negative or zero refractive index, extreme permittivity or permeability, etc.) with the great potential offered by $PT$-symmetry. An important class of metamaterials which are highly unexplored under the concept of $PT$-symmetry are the so-called chiral metamaterials [26], i.e. metamaterials composed of building-blocks that cannot be superimposed with their mirror images using translations and rotations [27,35]. A chiral medium is characterized by strong magneto-electric coupling (i.e. an applied electric field results not only to electric but also to magnetic polarization; same for an applied magnetic field). This results to different index of refraction for left-handed circularly polarized (LCP) waves and right-handed circularly polarized (RCP) waves, while circularly polarized waves are the eigensolutions of the wave equation in such media. We have to note here that chirality appears not only in metamaterials, but in some naturally occurring materials as well [28]. However, its effects in metamaterials are huge in comparison with those in natural materials, because in the latter, as opposed to the former, the structural features and the corresponding currents responsible for chirality are restricted by atomic size.

Among the interesting and particularly useful effects associated with chiral metamaterials are (a) the so called circular dichroism (i.e. different absorption for LCP and RCP waves), resulting, e.g. to a transformation of an incident linearly polarized wave to elliptical, and (b) optical activity, i.e. the rotation of the polarization of an initially linearly polarized wave passing through a chiral medium. As was mentioned before, these effects are quite strong in metamaterials even in thin layers of them. Indeed, extremely high optical activity and strong circular dichroism have been demonstrated in certain designs of chiral metamaterials, at frequencies from microwaves up to the visible range [29]. Moreover, negative refractive index has been predicted and demonstrated in such metamaterials [30-33].

The above mentioned phenomena and possibilities make the chiral metamaterials extremely valuable for a large variety of applications, especially applications based on the wave-polarization control, such as ultrathin circular polarizers and polarization rotators, wave-plates, polarization-modulators, etc. Based on the above, the idea to combine chiral media with $PT$-symmetry can give great promise for unique polarization-related effects, such as asymmetric polarization rotation, circularly polarized-wave lasers, etc. Indeed, first studies, investigating EM wave scattering by a two-dimensional chiral $PT$-symmetric system under normal incidence [26] showed the independence of EP from the chirality parameter (i.e. the parameter quantifying the magnetoelectric coupling) and the possibility to control separately and superimpose almost at will the $PT$-related features (e.g. EPs, CPA-laser points) and the chirality-related features (ellipticity, polarization rotation), achieving, e.g., CPA-lasing of elliptically or circularly polarized waves.

The aim of the present work is to go beyond the existing effectively one-dimensional study, and to investigate $PT$-symmetric chiral metamaterials under oblique incidence, and explore the associated transmission and scattering features and possibilities, including the different attained $PT$-related phases. Assuming a simple geometry containing two homogeneous and isotropic chiral layers one with gain and one with loss, we start our investigation by deriving the necessary conditions for a chiral system to be $PT$-symmetric (Section II). Applying those conditions in our system, we find that the chirality has strong influence on $PT$-symmetric and broken phases (Section III). In particular, by tuning the chirality parameter and/or the angle of incidence we observe three distinct phases exhibited by our system: full $PT$-symmetry phase, mixed $PT$-symmetry phase and broken $PT$-symmetry phase. Furthermore, we observe asymmetric transmission effects, such as asymmetric ellipticity and asymmetric polarization rotation of the transmitted waves. We show that these asymmetric effects are not only chirality dependent but also incidence-angle dependent, a property giving great ability for an external control. All the above effects demonstrate the large potential of the $PT$-symmetric chiral systems in the control of the propagation and polarization properties of electromagnetic (EM) waves.

## II. PHYSICAL SYSTEM AND MAIN EQUATIONS

### A. *PT*-symmetry conditions in general chiral systems

The standard formulations for $PT$-symmetric optical systems rely on their analogy to a Schrödinger problem. Likewise, here, in order to derive the corresponding $PT$-symmetry conditions for chiral systems, we write Maxwell's equations as an eigenvalue problem. We start from the curl equations, $\nabla\times\boldsymbol{E}=i\omega\boldsymbol{B}$ and $\nabla\times\boldsymbol{H}=-i\omega\boldsymbol{D}$, in which we insert the chiral constitutive relations $\boldsymbol{D}=\varepsilon\varepsilon_0\boldsymbol{E}+i(\kappa/c)\boldsymbol{H}$ and $\boldsymbol{B}=\mu\mu_0\boldsymbol{H}-i(\kappa/c)\boldsymbol{E}$, where $\varepsilon,\mu,\kappa$ refer to the relative permittivity, permeability and the chirality parameter respectively ($\varepsilon_0,\mu_0$ are the vacuum permittivity and permeability respectively and $c$ the light velocity). Solving for the fields $\boldsymbol{B}$ and $\boldsymbol{D}$ we obtain an eigenvalue system of the following form:

$$\hat{A}\begin{bmatrix}\boldsymbol{B}\\ \boldsymbol{D}\end{bmatrix}=\omega\begin{bmatrix}\boldsymbol{B}\\ \boldsymbol{D}\end{bmatrix} \quad (1)$$

where

$$\hat{A}=\begin{bmatrix}-i\nabla A_1(\boldsymbol{r})\times -iA_1(\mathbf{r})\boldsymbol{\nabla}\times & -i\nabla A_2(\boldsymbol{r})\times -iA_2(\mathbf{r})\boldsymbol{\nabla}\times\\ i\nabla A_3(\boldsymbol{r})\times +iA_3(\mathbf{r})\boldsymbol{\nabla}\times & i\nabla A_4(\boldsymbol{r})\times +iA_4(\mathbf{r})\boldsymbol{\nabla}\times\end{bmatrix} \quad (2)$$

and $A_1(\boldsymbol{r})=-i\kappa c/(\varepsilon\mu-\kappa^2)$, $A_2(\boldsymbol{r})=\mu\mu_0c^2/(\varepsilon\mu-\kappa^2)$, $A_3(\boldsymbol{r})=\varepsilon\varepsilon_0c^2/(\varepsilon\mu-\kappa^2)$ and $A_4(\boldsymbol{r})=i\kappa c/(\varepsilon\mu-\kappa^2)$, where for simplicity we have omitted the space dependence in $\varepsilon,\mu$ and $\kappa$. (Note that due to the magneto-electric coupling in the constitutive relations it is not possible to achieve a simple eigenvalue problem working with the fields $\boldsymbol{E}$ and $\boldsymbol{H}$.)

The eigenvalue problem (1) is formally analogous to the Schrödinger problem in Quantum Mechanics, with the tensor-operator $\hat{A}$ playing the role of a Hamiltonian. For a *PT*-symmetric "Hamiltonian" we should require that $[\hat{P}\hat{T}, \hat{A}] = 0$, i.e. $\hat{P}\hat{T}\hat{A} = \hat{A}\hat{P}\hat{T}$.

Taking into account that $\hat{P}\hat{T}(i\nabla \times) = i\nabla \times$ and following the procedure of Refs. [18,26], we obtain the necessary conditions for a chiral system to be *PT*-symmetric, as follows:

$$\varepsilon(\boldsymbol{r}) = \varepsilon^*(-\boldsymbol{r}), \mu(\boldsymbol{r}) = \mu^*(-\boldsymbol{r}), \kappa(\boldsymbol{r}) = -\kappa^*(-\boldsymbol{r}) \qquad (3)$$

The achievement of full *PT*-symmetry phase though, i.e. of real eigenvalues, is obtained if, besides the "Hamiltonian", its eigenstates also should be *PT*-symmetric. In our system [see Eq. (1)], the role of eigenstates is played by the "vector" composed of the magnetic induction, $\boldsymbol{B}$, and the displacement field, $\boldsymbol{D}$. Since, $\boldsymbol{B}$ and $\boldsymbol{D}$ are related to $\boldsymbol{E}$ and $\boldsymbol{H}$ through linear chiral constitutive relations and because $\boldsymbol{E}$ and $\boldsymbol{H}$ are more commonly used in EM wave propagation description, we will examine/discuss in the following first the *PT*-symmetry properties of $\boldsymbol{E}$ and $\boldsymbol{H}$. Having in mind our model system shown in Fig. 1, we consider circularly polarized eigenwaves $\boldsymbol{E}_{\pm}$ [29,35] (the subscript '+' corresponds to RCP electromagnetic waves while the '-' to LCP waves) propagating, after their refraction at an air – chiral interface, along the *x-z* plane (see Fig. 1) (with incident wavevectors forming an angle $\theta_i$ with the *z*-axis); these are waves of the form $\boldsymbol{E}_{\pm} = E_0(\cos\theta_{\pm}\,\hat{x} \pm i\hat{y} + \sin\theta_{\pm}\,\hat{z})\; e^{i(q_{\pm}-\omega t)}$ where $q_{\pm} = k_{\pm}(z\cos\theta_{\pm} - x\sin\theta_{\pm})$, $E_0$ is the amplitude being a real number, $k_{\pm} = \omega(\sqrt{\varepsilon\mu} \pm \kappa)/c$ are the wave vectors in the chiral material (their space dependence is omitted here, for simplicity) and $\theta_{\pm}$ are the wave propagation/refraction angles. (The corresponding magnetic field can be found by Maxwell's equations as $\boldsymbol{H}_{\pm} = \mp i\sqrt{\varepsilon\varepsilon_0/\mu\mu_0}E_0(\cos\theta_{\pm}\,\hat{x} \pm i\hat{y} + \sin\theta_{\pm}\,\hat{z})e^{i(q_{\pm}-\omega t)}$.)

The time-reversal symmetry can be understood as equivalent to "motion reversal" of the process [36-39]. Therefore, the action of time- reversal operator, $\hat{T}$ $(i \to -i)$, in $\boldsymbol{E}_{\pm}$ and $\boldsymbol{H}_{\pm}$ gives

$$\hat{T}\begin{pmatrix}\boldsymbol{E}_{\pm}\\ \boldsymbol{H}_{\pm}\end{pmatrix} = \begin{pmatrix} E_0(\cos\theta_{\pm}^*\hat{x} \mp i\hat{y} + \sin\theta_{\pm}^*\hat{z})e^{-i(q_{\pm}^*+\omega t)} \\ \pm i\sqrt{\varepsilon^*\varepsilon_0/\mu^*\mu_0}E_0(\cos\theta_{\pm}^*\hat{x} \mp i\hat{y} + \sin\theta_{\pm}^*\hat{z})e^{-i(q_{\pm}^*+\omega t)} \end{pmatrix} \qquad (4a)$$

The action of parity, $\hat{P}$, operator flips the space $(\hat{x} \to -\hat{x}, \hat{y} \to -\hat{y}, \hat{z} \to -\hat{z})$, i.e.

$$\hat{P}\hat{T}\begin{pmatrix}\boldsymbol{E}_{\pm}\\ \boldsymbol{H}_{\pm}\end{pmatrix} = \begin{pmatrix} -E_0(\cos\theta_{\pm}^*\hat{x} \mp i\hat{y} + \sin\theta_{\pm}^*\hat{z})e^{i(q_{\pm}^*-\omega t)} \\ \mp i\sqrt{\varepsilon^*\varepsilon_0/\mu^*\mu_0}E_0(\cos\theta_{\pm}^*\hat{x} \mp i\hat{y} + \sin\theta_{\pm}^*\hat{z})e^{i(q_{\pm}^*-\omega t)} \end{pmatrix} \qquad (4b)$$

with the star denoting complex conjugation. It is also noted that, the complex conjugation in the propagating angles $\theta_{\pm}$ comes from Snell's law [27,34]. Employing the conditions (3) to associate the quantities $\varepsilon, \mu, q_{\pm}$ and $\theta_{\pm}$ with their PT-counterparts, we conclude that

$$\hat{P}\hat{T}\begin{pmatrix}\boldsymbol{E}_{\pm}\\ \boldsymbol{H}_{\pm}\end{pmatrix} = -\begin{pmatrix}\boldsymbol{E}_{\mp}\\ \boldsymbol{H}_{\mp}\end{pmatrix}. \qquad (5)$$

From Eq. (5) one can see that under the action of *PT* an initial RCP wave transforms to LCP and vice versa (except a phase factor). Considering this and the constitutive relations for chiral media in connection with Eqs. (3), one can derive the same conclusion for the $\boldsymbol{B}$ and $\boldsymbol{D}$ fields; this implies that the eigenvectors in our case are not fully *PT*-symmetric. This non full *PT*-symmetry of the eigenvectors though does not forbid the existence of real eigenvalues in our problem. Here the reality of the eigenvalues (*PT*-symmetric phase) is ensured by the degeneracy of RCP/LCP eigenwaves with respect to the frequency [26]; both RCP and LCP waves share a common degenerate eigenvalue ω, as is concluded from Eqs. (1) and (2). Therefore, the conditions (3) ensure both the *PT*-symmetry of the "Hamiltonian" and the possibility of existence of real eigenvalues, and thus of a full PT-symmetric phase in a chiral system. In the following, we will exploit this possibility and investigate further its consequences in scattering configurations, where the "role" of the Hamiltonian is undertaken by the scattering matrix [15] involving all the transmission and reflection elements for the systems under consideration.

### B. Scattering of a linearly polarized plane wave by a *PT*-symmetric chiral system

In this section we discuss the calculation of the transmission and reflection properties of a *PT*-symmetric chiral system, i.e. of a system obeying the conditions (3), for a linearly polarized incident plane wave under oblique incidence. Our model system is composed of two chiral slabs, infinite along *x*- and *y*-directions and of thickness $d$ along *z*, as shown in Fig.1. The structure has a total thickness $L = 2d$ and is imbedded in air. The first slab is confined between $-L/2 \leq z \leq 0$ and the second slab is confined between $0 \leq z \leq L/2$. The system is characterized by complex material parameters, which are the permittivity, permeability and chirality of each slab.

We assume that we have either TM or TE polarized incident plane waves, arriving at angle $\theta_i$ from the left side, as shown in Fig. 1:

$$\text{TM:}\begin{cases} \boldsymbol{E}_i^{(A)} = E_{i\parallel}(cos\theta_i\hat{x} + sin\theta_i\hat{z})\; e^{ik_i(z\,cos\theta_i - x\,sin\theta_i)} \\ \boldsymbol{H}_i^{(A)} = \frac{E_{i\parallel}}{Z_0}\hat{y}\; e^{ik_i(z\,cos\theta_i - x\,sin\theta_i)} \end{cases} \qquad (6)$$

and

$$\text{TE:}\begin{cases} \boldsymbol{E}_i^{(A)} = E_{i\perp}\hat{y}\; e^{ik_i(z\,cos\theta_i - x\,sin\theta_i)} \\ \boldsymbol{H}_i^{(A)} = -\frac{E_{i\perp}}{Z_0}\;(cos\theta_i\hat{x} + sin\theta_i\hat{z})e^{ik_i(z\,cos\theta_i - x\,sin\theta_i)} \end{cases} \qquad (7)$$

where $E_{i\parallel}, E_{i\perp}$ are the amplitudes for TM and TE waves, respectively, $k_i = \omega/c$ is the vacuum wavenumber and $Z_0 = \sqrt{\mu_0/\varepsilon_0}$ is the vacuum wave impedance. The subscript ∥ in the fields indicates the component that lies in the plane of incidence (*x-z* in Fig. 1) while the subscript ⊥ indicates the perpendicular component.

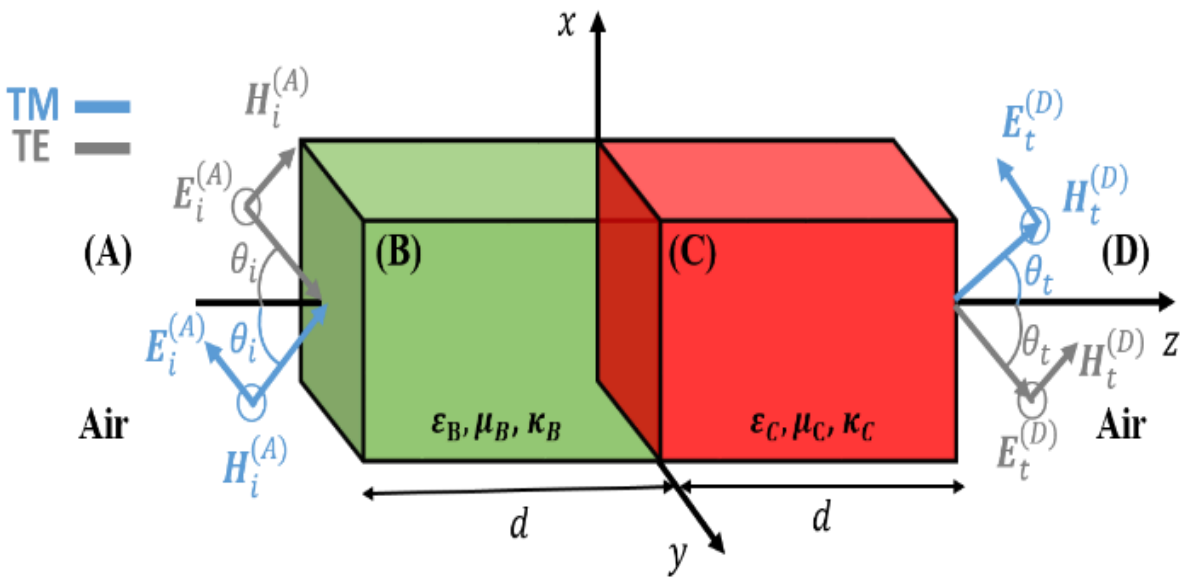

Figure 1. The model *PT*-symmetric chiral system studied in the present work. It consists of two layers characterized by complex permittivities, permeabilities and chiralities. It is illuminated by obliquely incident either TM-polarized plane wave (blue) or TE-polarized plane wave (grey).

Writing the fields in the chiral slabs as a linear combination of RCP and LCP waves and applying the boundary conditions of the continuity of the tangential components of the electric and magnetic fields at the three interfaces $z = -d, z = 0$ and $z = d$, one can calculate the fields inside and outside the chiral slabs through the solution of a 12x12 linear system of equations (see Appendix A for details); the system is solved numerically due to its complex expressions. From the solution of the 12x12 system one can obtain the reflection and transmission amplitudes for TM and TE polarized electromagnetic waves incident on the chiral *PT*-symmetric structure, which are defined as

$$\begin{pmatrix} E_{t\perp} \\ E_{t\parallel} \end{pmatrix} = \begin{pmatrix} t_{\perp\perp} & t_{\perp\parallel} \\ t_{\parallel\perp} & t_{\parallel\parallel} \end{pmatrix} \begin{pmatrix} E_{i\perp} \\ E_{i\parallel} \end{pmatrix} = T_{lin} \begin{pmatrix} E_{i\perp} \\ E_{i\parallel} \end{pmatrix}, \quad (8)$$

and

$$\begin{pmatrix} E_{r\perp} \\ E_{r\parallel} \end{pmatrix} = \begin{pmatrix} r_{\perp\perp} & r_{\perp\parallel} \\ r_{\parallel\perp} & r_{\parallel\parallel} \end{pmatrix} \begin{pmatrix} E_{i\perp} \\ E_{i\parallel} \end{pmatrix} = R_{lin} \begin{pmatrix} E_{i\perp} \\ E_{i\parallel} \end{pmatrix}. \quad (9)$$

The subscript $t$ in (8) indicates the transmitted wave while the subscript $r$ in (9) the reflected wave. The first subscript in the amplitudes $r$ and $t$ indicates the output wave polarization while the second the incident wave polarization.

In order to fully characterize the polarization state of the waves passing through our chiral system, we can calculate the ellipticity, $\chi$ (directly connected to the circular dichroism), and the rotation angle, $\psi$ (i.e. orientation angle of the polarization ellipse in respect to the axis $y$, a measure of the optical activity), for those waves. This can be done by calculating the corresponding Stokes parameters [34], which describe completely the state of polarization of a wave and are directly connected with $\tan(\chi)$ and $\tan(2\psi)$, as is discussed in detail in Appendix B.

## C. Circularly polarized wave incidence and the scattering matrix

As has been already mentioned, in scattering configurations the identification of the different *PT*-related phases and the exceptional points can be done by examining the scattering matrix eigenvalues. In the *PT*-symmetric phase the eigenvalues, $\sigma_i$, of the scattering matrix are unimodular (i.e. obey $|\sigma_i| = 1$), in contrast to the *PT*-broken phase in which $|\sigma_i| \neq 1$ and the eigenvalues form pairs of reciprocal magnitude [7,13] (here the subscript $i$ counts the different eigenvalues). At the symmetry breaking point, i.e. the exceptional point, two or more eigenvalues coincide.

To investigate the different *PT*-related phases in our system and to identify the position of exceptional point/points the first step is to calculate the scattering matrix, $S$, and its eigenvalues. Taking into account that the eigenwaves in chiral systems are the circularly polarized waves, it can be seen that the elements of the $S$ matrix can be obtained from the reflection and transmission coefficients presented in the previous subsection after transformation in a circular polarization basis (see Appendix A). To identify though the proper scattering matrix formulation (i.e. the proper arrangement of the transmission and reflection coefficients within the scattering matrix) it is essential to write both the incoming to the system and the outgoing waves in circular polarization basis and illustrate the connection of the scattering matrix to the different input and output wave amplitudes. In this respect the electric field outside the *PT*-symmetric optical system can be expressed as

$$\boldsymbol{E}(z,t) = \begin{cases} B_1^+\hat{\mathbf{e}}_+ e^{iq} + B_1^-\hat{\mathbf{e}}_- e^{iq} + A_1^+\hat{\mathbf{e}}_+ e^{-iq} + A_1^-\hat{\mathbf{e}}_- e^{-iq} & , z < -d \\ B_2^+\hat{\mathbf{e}}_+ e^{-iq} + B_2^-\hat{\mathbf{e}}_- e^{-iq} + A_2^+\hat{\mathbf{e}}_+ e^{iq} + A_2^-\hat{\mathbf{e}}_- e^{iq} & , z > d \end{cases} \quad (10)$$

where $A_1^\pm, A_2^\pm, B_1^\pm$ and $B_2^\pm$ are amplitudes of the ingoing and outgoing RCP (+) and LCP (-) waves as shown in Fig. 2, $\hat{\mathbf{e}}_+ = cos\theta_i\hat{x} + i\hat{y} + sin\theta_i\hat{z}$, $\hat{\mathbf{e}}_- = cos\theta_i\hat{x} - i\hat{y} + sin\theta_i\hat{z}$ and $q = k_i(z\cos\theta_i - x\sin\theta_i)$, with $\theta_i$ the incidence angle. Therefore the system can be described by four input and four output ports (see Fig. 2), giving a 4x4 scattering matrix, $S$ (consisting of eight transmission and eight reflection coefficients), defined by

$$\begin{bmatrix} A_1^- \\ B_2^+ \\ A_1^+ \\ B_2^- \end{bmatrix} = S \begin{bmatrix} B_1^+ \\ A_2^- \\ B_1^- \\ A_2^+ \end{bmatrix} \equiv \begin{bmatrix} r_{-+}^{left} & t_{--}^{right} & r_{--}^{left} & t_{-+}^{right} \\ t_{++}^{left} & r_{+-}^{right} & t_{+-}^{left} & r_{++}^{right} \\ r_{++}^{left} & t_{+-}^{right} & r_{+-}^{left} & t_{++}^{right} \\ t_{-+}^{left} & r_{--}^{right} & t_{--}^{left} & r_{-+}^{right} \end{bmatrix} \begin{bmatrix} B_1^+ \\ A_2^- \\ B_1^- \\ A_2^+ \end{bmatrix}. \quad (11)$$

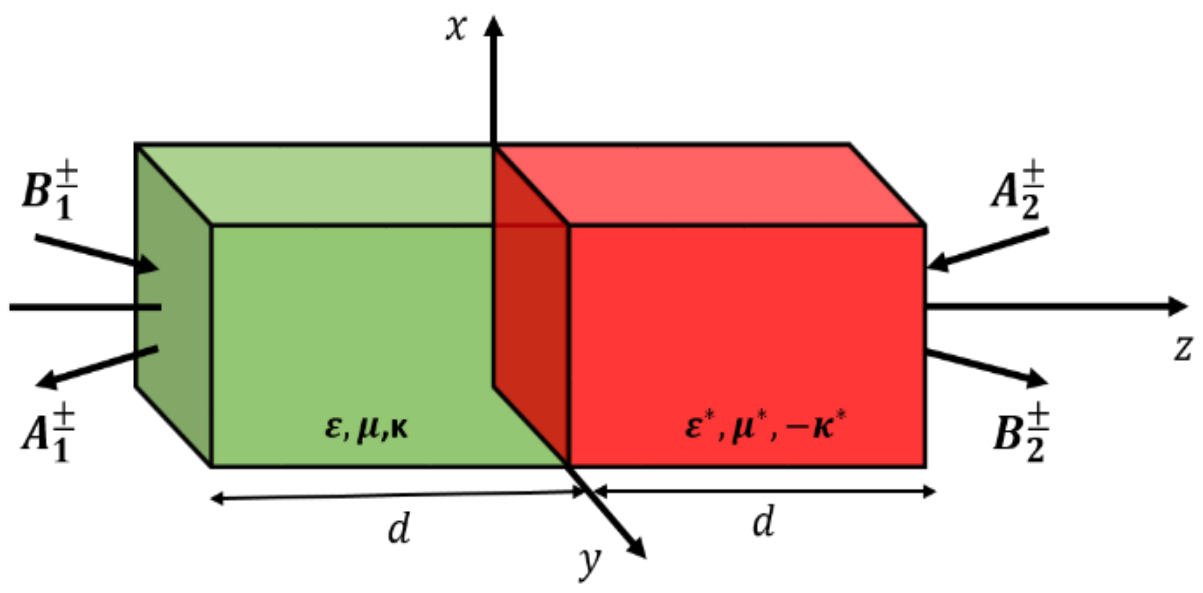

Figure 2. Schematic representation of the scattering of circularly polarized plane waves by a two-layer *PT*-symmetric chiral optical system.

In Eq. (11) $t_{++}^{left}, t_{--}^{left}, t_{+-}^{left}, t_{-+}^{left}$ and $r_{++}^{left}, r_{--}^{left}, r_{+-}^{left}, r_{-+}^{left}$ are the transmission and reflection coefficients for LCP(-)/RCP(+) light incident from the left while

$t_{++}^{right}$, $t_{--}^{right}$, $t_{+-}^{right}$, $t_{-+}^{right}$ and $r_{++}^{right}$, $r_{--}^{right}$, $r_{+-}^{right}$, $r_{-+}^{right}$ are the transmission and reflection coefficients for LCP (-) /RCP (+) light incident from the right (for their calculation see Appendix A). As was mentioned also in the previous paragraph, depending on the arrangement of the amplitudes $A_1^{\pm}, A_2^{\pm}, B_1^{\pm}, B_2^{\pm}$ the scattering matrix $S$ can be formulated in several ways. However, the $S$-matrix defined as in Eq. (11) satisfies the relation $PTS(\omega^*)PT = S^{-1}(\omega)$ [15], which is a fundamental condition obeyed by $PT$-symmetry.

## III. REPRESENTATIVE RESULTS

### A. Controlling *PT*-symmetry phase

In this subsection, we investigate the different possible attainable phases in our double-layer $PT$-symmetric structure shown in Fig 2, by numerically evaluating the scattering matrix eigenvalues under oblique incidence. We assume first the simplest case of chirality $\kappa = 0$ and calculate the eigenvalues of the scattering matrix for TE and TM electromagnetic waves, separately. (Note that in this case the $S$ matrices are 2x2 and defined as in [13, 22].) Results for the eigenvalues vs frequency are shown in Figs. 3(a) and 3(b). A surprising property is that the EP frequencies for the TM and TE polarizations *under oblique incidence* are different [25]; this implies the existence of an intermediate phase, to be referred to as mixed phase, where the $PT$-symmetry phase still exists for one polarization while it is broken for the other in the same system and incident wave direction.

The mixed phase is accessible by employing circularly polarized (CP) incident waves (since they can be written as a linear combination of TM and TE waves). Indeed, calculating the eigenvalues of the scattering matrix for CP waves [Eq. (13)], as a function of frequency, we observed the three different possible phases, as shown in Fig. 3(c). In the full $PT$ symmetric phase all the eigenvalues of the scattering matrix maintain their unimodular nature, in the mixed $PT$ phase (light grey in Fig. 3 (c)) a pair of the eigenvalues is unimodular while the other is not and in the fully broken $PT$ phase (dark grey) all the eigenvalues are non-unimodular.

Besides the existence of three different phases, and thus two EPs in our system, the other particularly interesting and peculiar feature is that the position of these two EPs and thus the boundaries of the different phases can be highly controlled by the angle of incidence. This is demonstrated in Fig. 3(d), where the dependence of the mixed phase extent as a function of the angle of incidence from $0^o$ to $90^o$ degrees is shown. One can see that the frequencies of the two EPs fend off as we increase the incidence angle, while the two EPs, as is expected, coincide for normal incidence.

To investigate the impact of the chirality on the position of EPs and the different phases of $PT$- symmetric media we consider a chiral $PT$-symmetric bilayer as illustrated in Fig. 2; the first layer (gain) has permittivity $\varepsilon(-z) = \varepsilon_r - \varepsilon_i i$, permeability $\mu(-z) = \mu_r - \mu_i i$ and chirality $\kappa(-z) = -\kappa_r + \kappa_i i$ while the second layer (loss) $\varepsilon(z) = \varepsilon_r + \varepsilon_i i$, $\mu(z) = \mu_r + \mu_i i$ and chirality parameter $\kappa(z) = \kappa_r + \kappa_i i$, respectively, so as to fulfill the necessary conditions for $PT$-symmetry [Eq. (3)]. Considering permittivity and permeability values the same as in Fig. 3, introducing a non-zero chirality and repeating the calculations of the eigenvalues of the scattering matrix, we obtain what is shown in Fig. 4(a). ). In Fig. 4(a) one can see that when the chirality parameter is relatively weak, the same three phases shown in Fig. 3(c) are obtained, but with slightly different boundaries.

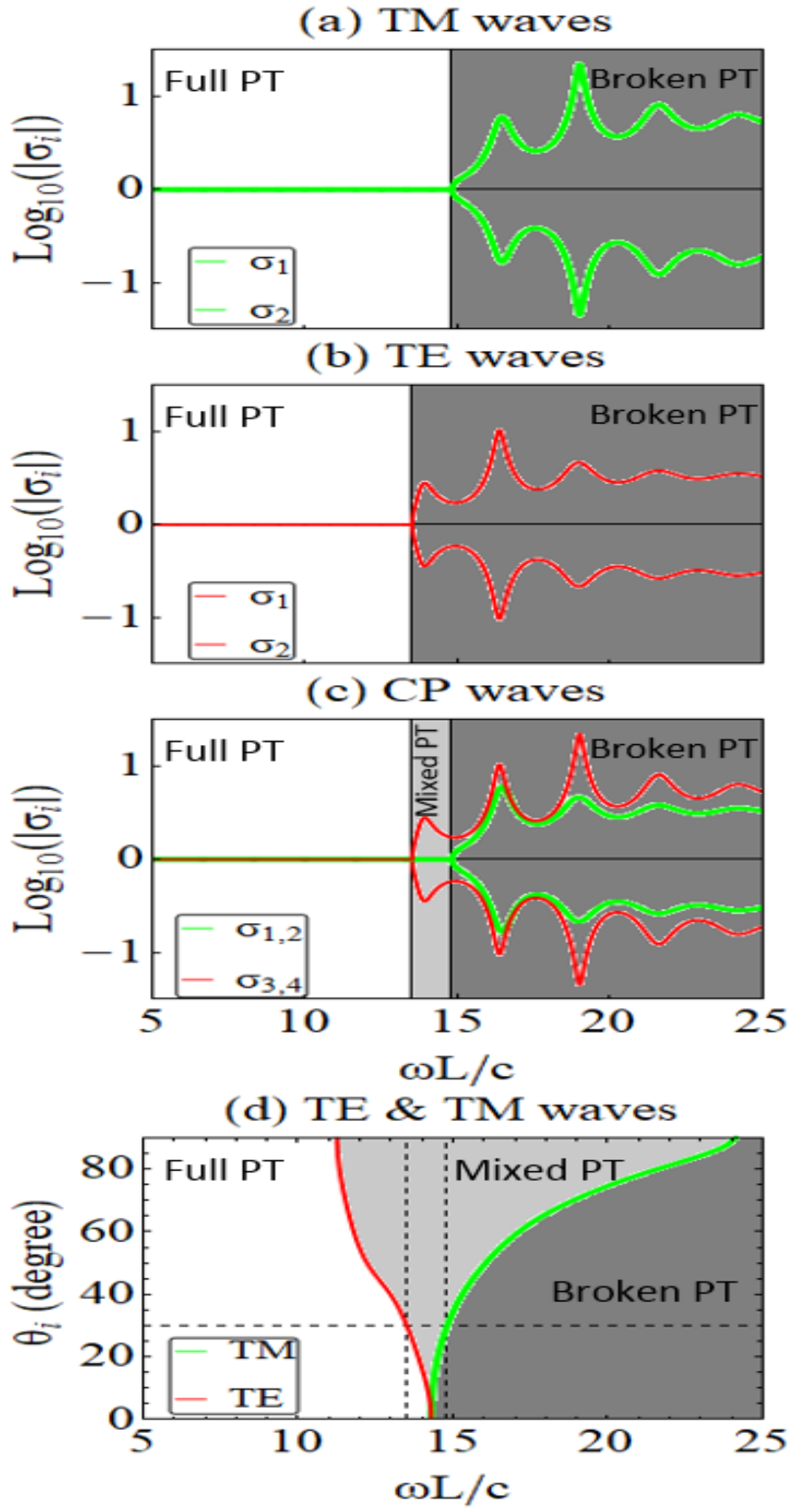

Figure 3. $PT$-symmetric (white), mixed $PT$-symmetric (light-gray) and broken $PT$-symmetric (dark-gray) phases for obliquely incident waves for the system shown in Fig. 1 with chirality $\kappa = 0$. The figure shows the eigenvalues of scattering matrices at $\theta_i = 30^o$ : (a) for TM, (b) TE and (c) for CP waves,. The exceptional points versus frequency as the incidence angle varies are illustrated in (d). The relevant material parameters of the two media are: $\varepsilon(-z) = 4 - 0.6i$, $\varepsilon(z) = 4 + 0.6i$, $\mu(-z) = 1.5 - 0.15i$, $\mu(z) = 1.5 + 0.15i$ .

*This clearly shows that one important impact of chirality is the tuning of the different phases and the EPs.* We have to note here that this is in contrast to what has been observed for normal incidence, where the position of EP is totally independent of chirality [26]. To investigate in more detail the impact of chirality on the different phases and the EPs of our system we calculated the scattering matrix eigenvalues for three distinct frequencies, corresponding to full *PT*, mixed *PT* and broken *PT* phases in Fig. 4(a) (marked by the colored dots) as a function of the chirality parameter. The result is shown in Figs. 4(b-d), where by changing the chirality under *oblique incidence* a quite unexpected behavior is observed.

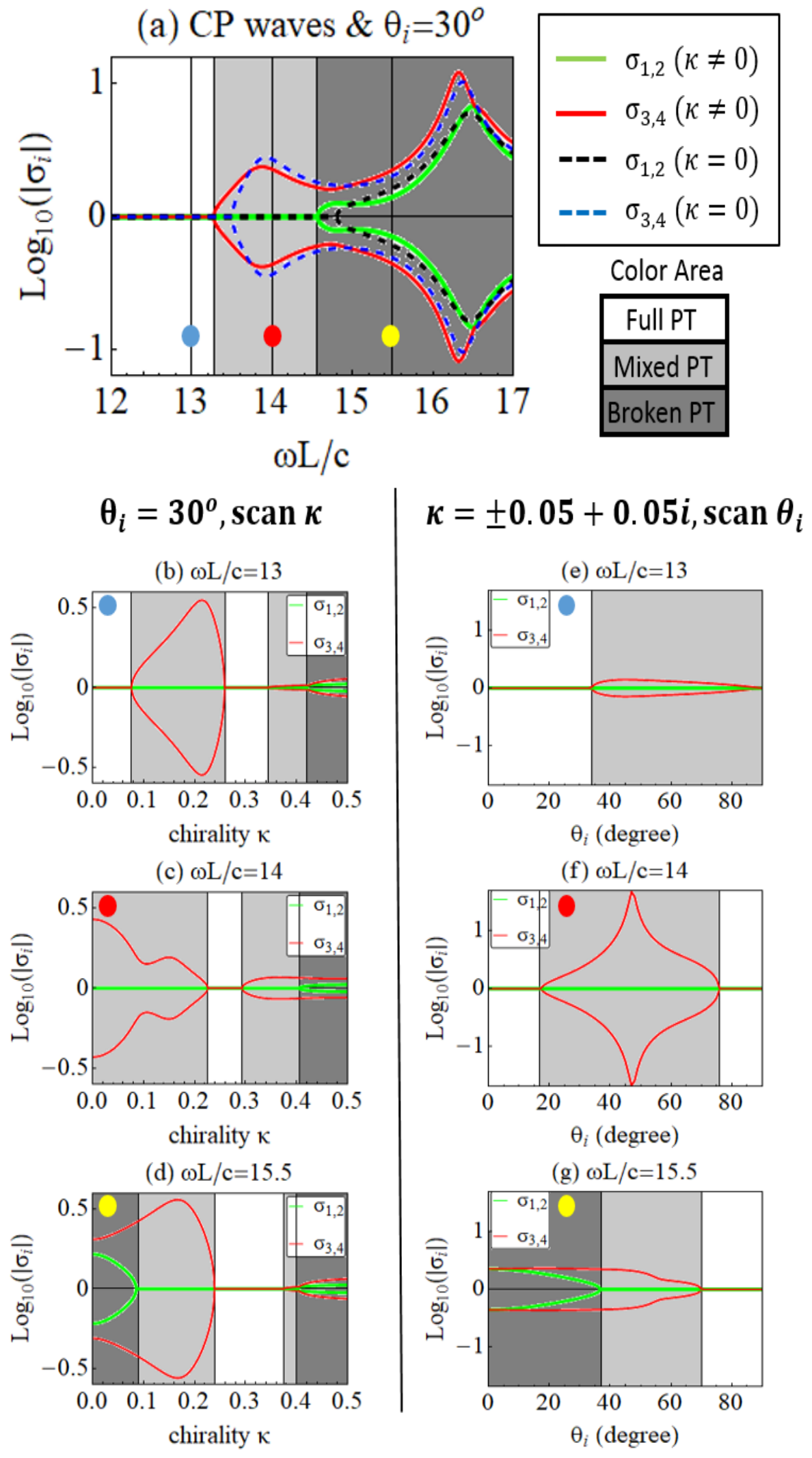

Figure 4. *PT*-symmetric, mixed *PT*-symmetric and broken *PT*-symmetric phases in a chiral *PT*-system under oblique incidence of CP waves. The relevant material parameter of the two media are: $\varepsilon(-z) = 4 - 0.6i$, $\varepsilon(z) = 4 + 0.6i$, $\mu(-z) = 1.5 - 0.15i$, $\mu(z) = 1.5 + 0.15i$ and $\kappa(-z) = -0.05 + 0.05i$, $\kappa(z) = 0.05 + 0.05i$. Panel (a) shows the eigenvalues, $|\sigma_i|, i = 1 - 4$, of $S$ matrix vs frequency for $\theta_i = 30^o$. Panels (b), (c) and (d) show these eigenvalues as a function of chirality parameter, while panels (e), (f) and (g) show them as a function of incidence angle for three different dimensionless frequencies, $\frac{\omega L}{c} = 13.0$, 14.0 and 15.5, respectively.

By increasing the chirality parameter, both real and imaginary parts, the system passes from *PT*-symmetric phase to mixed *PT* phase (light grey), then re-enters to *PT*-symmetric phase and with further increase of chirality re-enters to mixed *PT* phase and, finally, ends up in the fully broken phase. Analogous behavior is shown in Figs. 4(c) and 4(d), *demonstrating that by changing the chirality not only can we tune the EPs but we can also access different EPs and observe different phases and phase re-entries.* Note that it is possible to achieve analogous control of the *PT*-symmetry-related phase by changing only the real or only the imaginary part of the chirality - see Appendix C for details; this give us an additional degree of freedom for *PT*-phases control.

Another important possibility regarding the control of the different *PT*-phases and the exceptional points is observed if one calculates the eigenvalues of the $S$ matrix as the incidence angle varies. Performing such calculations, for the same parameters as in Fig. 4(a) and for three characteristic frequencies we obtain the results shown in Figs. 4(e)-(g). *Figs. 4(e)-(g) demonstrate the possibility for external dynamic control of the different PT-symmetry-related phases and the exceptional points.* By changing the incidence angle, different phases are accessible, as well as phase re-entries, as, e.g. in Fig. 4(f). Therefore, our results show a very rich behavior of *PT*-symmetric chiral systems, including existence of mixed phases and phase re-entries; all these features can be highly controlled by tuning the chirality parameter and/or the angle of incidence, allowing thus a full external and even dynamic control of the *PT* system phases.

To investigate the impact of the different *PT*-symmetry-related phases on the transmission and reflection properties of our chiral system we examine next the reflection and transmission coefficients for obliquely incident circularly polarized waves as a function of chirality. We do that for the case of Fig. 4(b), i.e. $\omega L/c$=13 and at $\theta_i = 30^o$, where a rich behavior regarding different *PT*-phases and EPs has been observed. The result is presented in Figs. 5, where we plot the reflected and transmitted power coefficients for right circularly polarized (RCP) and left circularly polarized (LCP) waves incident from both sides of our system, as a function of chirality. A worth-noticing feature in Fig. 5, which is absent for normal incidence [26], is the presence of cross-polarized transmission and of co-polarized reflection coefficients. These cross-polarized terms (outcome of the breaking of the 4-fold rotational structure symmetry in the plane perpendicular to the propagation direction - due to the interfaces) seem to be enhanced in the mixed *PT* and the broken *PT* phases.

Another observation from Fig. 5 is the strong *asymmetry* in both the reflection and the cross-polarized transmission coefficients for waves impinging from the left and the right side of the system. Moreover, there is a strong asymmetry in the system response for RCP and LCP waves. In particular, the system seems to cause attenuation of any incident RCP wave or transformation of it to LCP wave (either through reflection or through transmission) while it enhances the LCP wave impinging on it. This LCP-RCP asymmetry is a direct consequence of the fact that we have chosen positive imaginary part for the chirality parameter (implying a specific structure circular dichroism response) and it is reversed for structures with negative imaginary part of κ.

The above-mentioned asymmetric effects reveal a very rich structure behaviour and a possibility of an at will control of the scattering and polarization of the waves impinging on our structure. This rich behaviour and the scattering control possibilities will become more evident in the next section where we investigate the scattering properties of the structure under a linearly polarized incident wave.

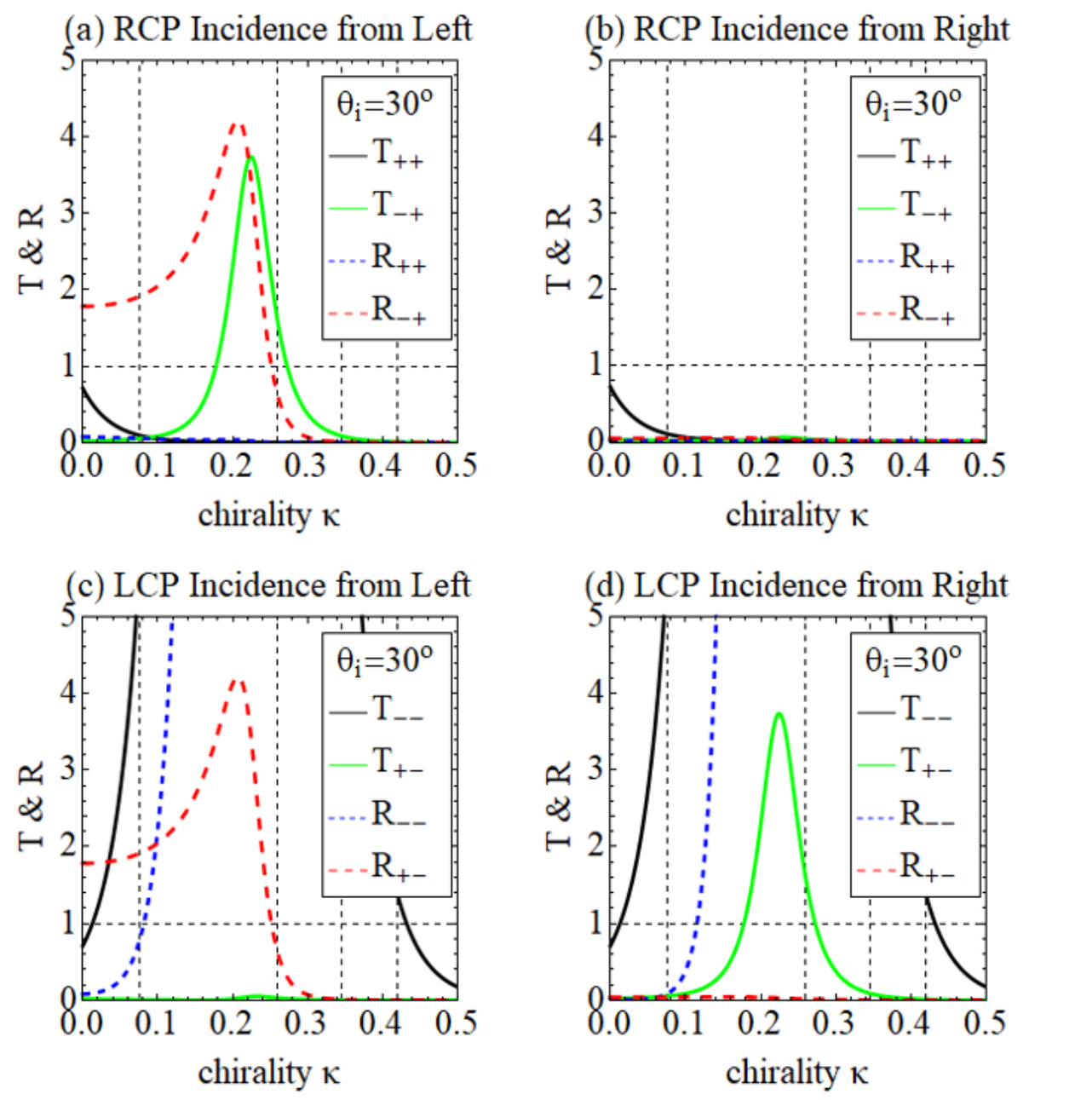

Figure 5. Reflected and transmitted power coefficients for the chiral *PT*-system shown in Figs. 1, 2 under oblique incidence of CP waves. The relevant material parameters of the two media are the same as in Fig. 4. Panels (a) and (b) show the reflection and transmission vs chirality for RCP incident waves. Panels (c) and (d) show the reflection and transmission vs chirality for LCP incident waves. The angle of incidence is 30 degrees and the dimensionless frequency is $\frac{\omega L}{c} = 13$. The positions of the exceptional points (see Fig. 4b) are marked by vertical dotted lines.

## B. Asymmetric effects in chiral *PT*-symmetric systems excited by a linearly polarized wave

As was mentioned above, excitation of a chiral system with a linearly polarized wave allows for a more detailed investigation of the polarization control properties and capabilities of the *PT*-symmetric chiral structure; it also allows for a more direct comparison with potential experiments. Here we consider a chiral *PT*-symmetric system as the one discussed in the previous subsection (i.e. with the same material parameters) and calculate the transmission coefficients $T_{ij} = |t_{ij}|^2$, with $t_{ij}$ the amplitudes defined in Eq. (8), for a linearly polarized wave incident from either the left or the right side of the system, as shown in Fig. 1. Fig. 6 shows these transmission coefficients for both normal and oblique incidence. It is evident that for normal incidence the transmission amplitudes for both the left- and right-side incident plane waves are exactly the same (see Figs. 6(a) and (b)). This is not the case though for oblique incidence, as can be seen comparing Figs. 6(c) and (d). For obliquely incident waves the transmission amplitudes are quite different for TM and TE incident waves and the cross-polarized coefficients are side-dependent. In particular, $T_{\perp\parallel}$ and $T_{\parallel\perp}$ interchange for opposite propagation directions ($T_{\perp\parallel}^{(left)} = T_{\parallel\perp}^{(right)}$ and $T_{\parallel\perp}^{(left)} = T_{\perp\parallel}^{(right)}$) (co-transmissions $T_{\parallel\parallel}$ and $T_{\perp\perp}$ are side-independent, as dictated by reciprocity). The side-dependence of the cross-polarized transmission terms implies *side-dependent transmitted wave polarization*.

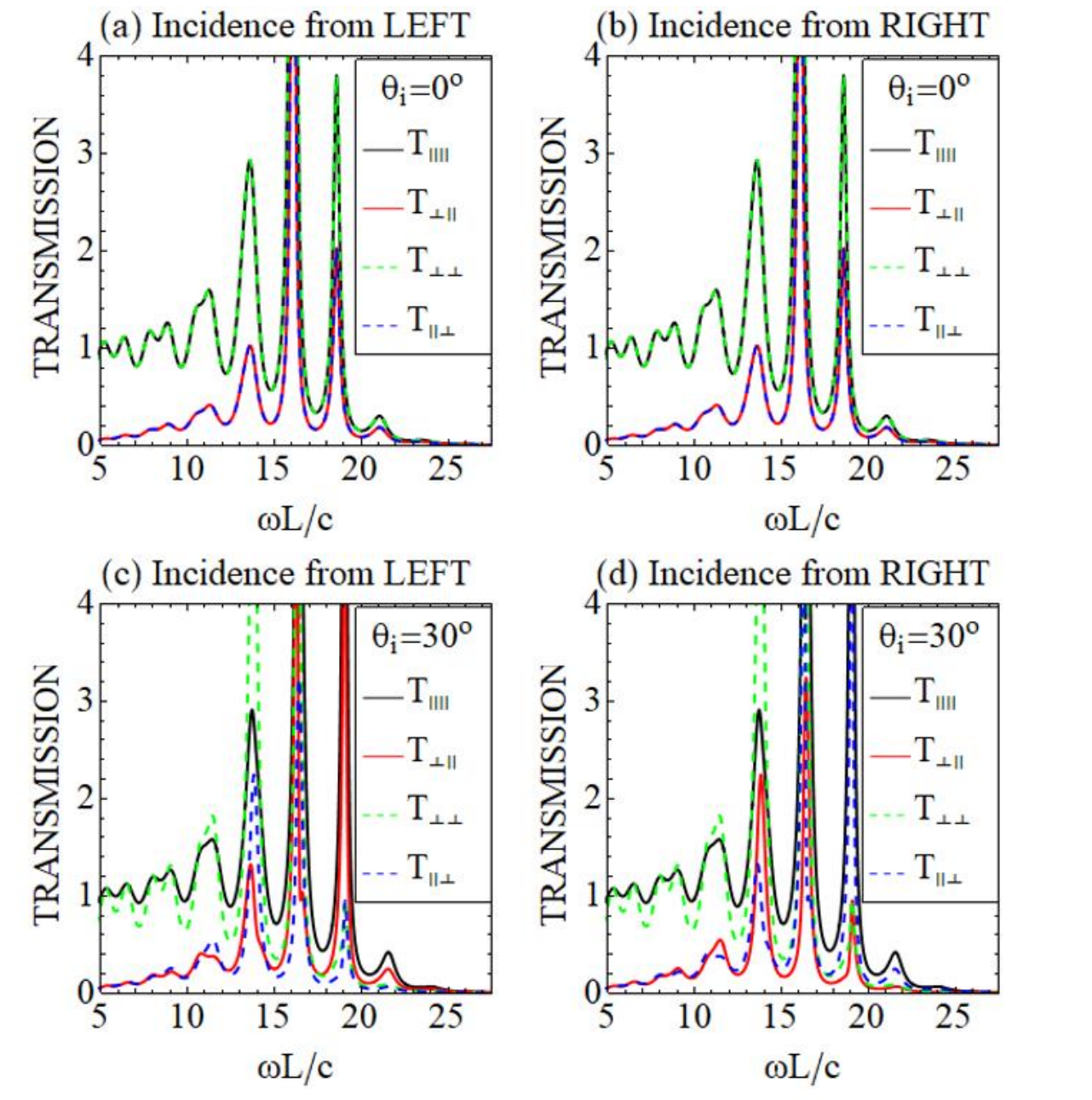

Figure 6. Transmission coefficients for TM and TE polarized plane waves incident from the left [(a),(c)] and right [(b),(d)] side of the PT-symmetric system shown in Fig. 1. The relevant material parameter of the two media are: $\varepsilon(-z) = 4 - 0.6i$, $\varepsilon(z) = 4 + 0.6i$, $\mu(-z) = 1.5 - 0.15i$, $\mu(z) = 1.5 + 0.15i$ and $\kappa(-z) = -0.05 + 0.05i$, $\kappa(z) = 0.05 + 0.05i$. For (c) and (d) the angle of incidence is 30 degrees, while for (a) and (b) $\theta_i = 0^o$.

To analyze further the polarization properties of the transmitted waves we calculated their ellipticity and optical activity (i.e. orientation angle of polarization ellipse) as a function of frequency for the transmission data presented in Fig. 6 (see Appendix B for the details of the calculation). For the case of normal incidence (Figs. 7(a) and (b)) both the ellipticity and optical activity are independent of the incident side, i.e. they are fully symmetric, as is expected given the fully symmetric transmission coefficients. (Due to the opposite sign of the Real(κ) in the two layers, imposed by Eq. (3), the optical rotation occurring in the first layer is cancelled out when the wave passes through the second layer, resulting to zero optical rotation – see Fig. 7(b).)

In the case of oblique incidence, since the transmittance for co-polarized and cross polarized waves are different for TE and TM waves and also side dependent, the ellipticity and the optical activity are also different for TM and TE polarized waves and also side dependent, as illustrated in Fig. 7 (c)-(f). The results of Fig. 7 offer a clear demonstration of the rich polarization control possibilities offered by the *PT*-symmetric chiral systems. Note that while, in principle, asymmetric chiral effects can be also observed in non *PT*-symmetric systems, with *PT*-symmetric systems we achieve advanced polarization control capabilities combined with all *PT*-related functionalities (exceptional points, *PT*-symmetric phases, CPA-laser points for circularly polarized waves, anisotropic transmission resonances, etc.).

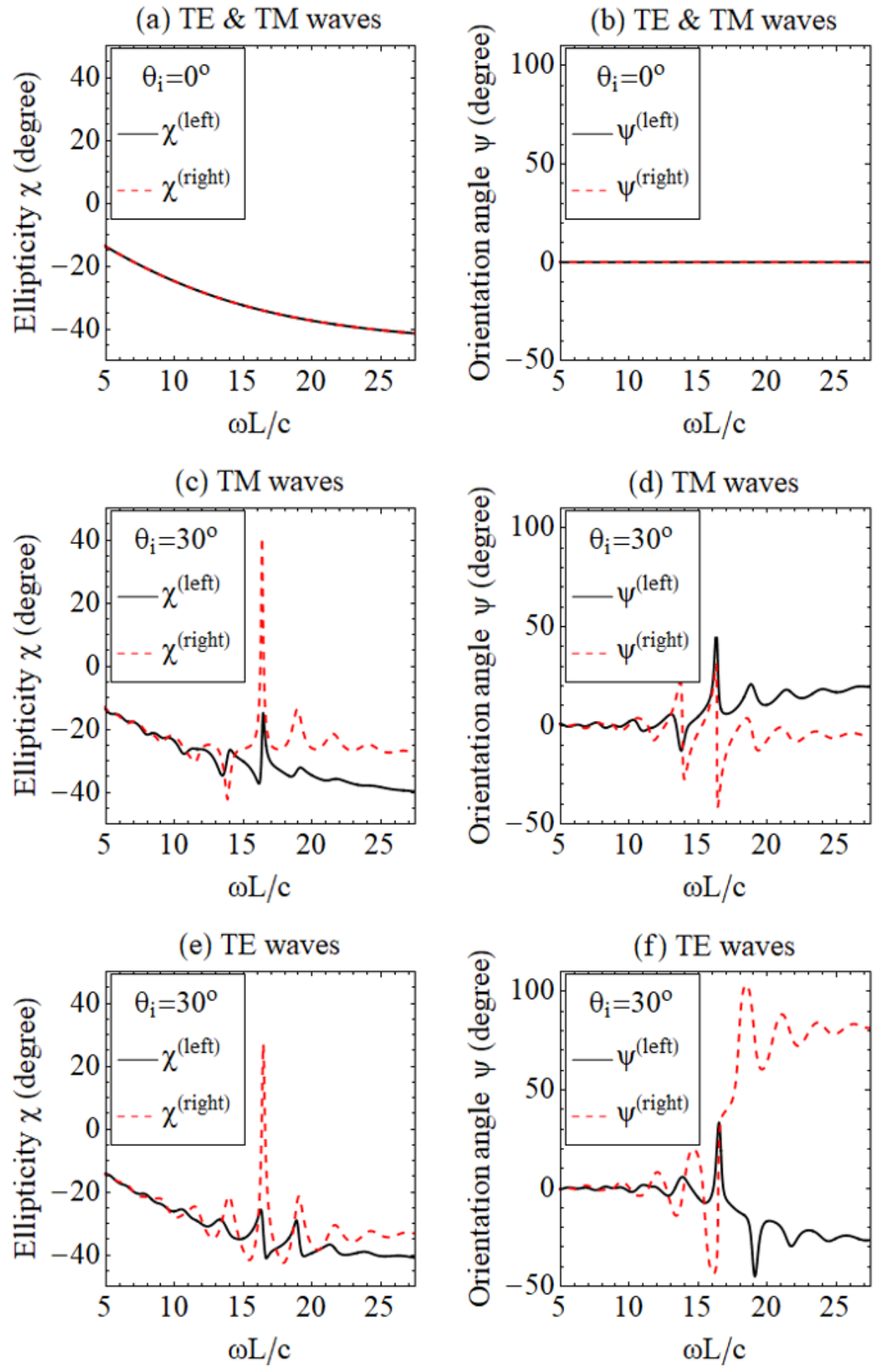

Figure 7. Ellipticity and orientation angle of polarization ellipse for TM [(c),(d)] and TE [(e),(f)] polarized plane waves incident from the left and right side of our *PT*-symmetric system shown in Fig. 1. The material parameter of the two media are the same as in Fig. 6.

To investigate further the polarization control possibilities offered by the *PT*-symmetric chiral systems, we calculate the transmission and reflection amplitudes with respect to the incidence angle, at dimensionless frequency $\frac{\omega L}{c} = 18.3$. The results are shown in Figs. 8(a)-8(d). Figs. 8(a) and (b) show the co-polarized (‖‖ or ⊥⊥) coefficients for TM and TE polarized incident waves, demonstrating a strong asymmetry (side-dependence) in the reflection coefficients which are also angle-dependent and thus angle-controllable, while the co-polarized transmissions are side-independent. Figs. 8(c), 8(d) show the cross-polarized coefficients, demonstrating side-asymmetric and angle-controllable cross-polarized transmissions. In particular, as mentioned also in the discussion of Fig. 6, the cross-polarized transmissions for TM and TE incident waves interchange for opposite propagation directions. An interesting feature that can be observed from Figs. 8(a) & (c) is that for TM incident waves we can have anisotropic transmission resonances (ATRs) [13,25], i.e. flux conserving unidirectional perfect transmission (here $T_{total}=1$, $R=0$ only for waves impinging from the right side of our system) at different incidence angles (marked with dashed lines in Figs. 8(a), 8(c)). This is also possible with TE waves (not shown here).

The angle dependence of the transmission coefficients shown in Figs. 8(a)-8(d) leads to angle-dependent transmitted wave ellipticities and optical activities, as is illustrated in Figs. 8(e)-8(h). Moreover, once again we see that the side-dependence of the cross polarized transmissions leads to asymmetric ellipticities and optical activities. The results of Figs. 8 demonstrate one more time the rich polarization behavior of the wave interacting with a *PT*-symmetric chiral system, and more importantly the possibility to highly control this polarization externally and even dynamically. The foreseen applications cover all the range of applications where passive or active polarization control is required, i.e. tunable polarization filters, tunable polarization isolators etc.

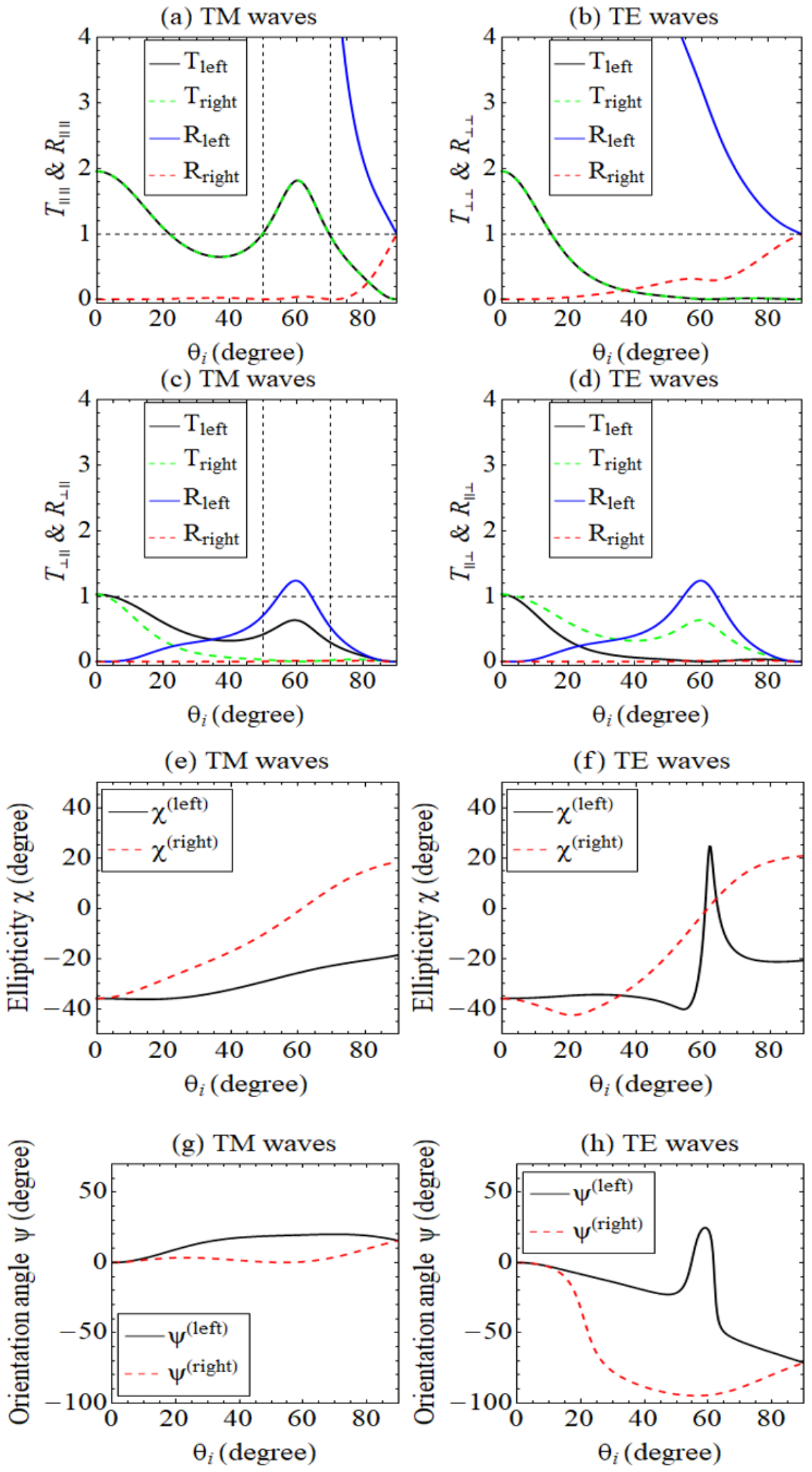

Figure 8. Co-polarized transmission and reflection coefficients [(a),(b)] and cross-polarized transmission and reflection [(c),(d)] for TM and TE incident waves as a function of incidence angle, at dimensionless frequency $\frac{\omega L}{c} = 18.3$. Panels (e)-(h) show the ellipticity and orientation angle of polarization ellipse for TM [(e),(g)] and TE [(f),(h)] polarized plane waves incident from the left and right side of our *PT*-symmetric system. The material parameters of the two media are the same as in Fig. 6.

# IV. CONCLUSION

In conclusion, we have investigated the influence of chirality in a *PT*-symmetric chiral structure under obliquely incident waves. We have shown that in addition to the fully *PT*-symmetric and the fully *PT*-broken phase for oblique incidence there is an intermediate phase, termed as "mixed", in which one of the TM, TE incident waves is in the *PT*-symmetric and the other in the *PT*-broken phase. As the *angle of incidence* and/or the *chirality vary*, a rich behavior of these phases and the associated exceptional points is exhibited, with unexpected re-entry phenomena controllable by both angle of incidence and chirality. Moreover, we have shown that combining *PT*-symmetry with chirality we can achieve novel propagation and scattering characteristics, like asymmetric (side-dependent) transmission, ellipticity and polarization rotation for a linearly polarized incident wave. This asymmetric behavior is angle dependent, offering an additional degree of freedom for controlling the transmission and the polarization properties of the electromagnetic waves interacting with *PT*-chiral structures. The aforementioned phenomena and the associated extensive opportunities for tuning EM waves are due to interplay between gain-loss and chirality in chiral *PT*-symmetric systems; these phenomena can be exploited in a large variety of applications, ranging from applications where advanced propagation and polarization control is required, to even sensing applications (exploiting the existence of multiple exceptional points, and the strong sensitivity associated with them [40-41]).

# ACKNOWLEDGMENTS

This work was supported by the Hellenic Foundation for Research and Innovation (HFRI) and the General Secretariat for Research and Technology (GSRT), under the HFRI PhD Fellowship grant (GA. no. 4820). As well as by the EU-Horizon2020 FET projects Ultrachiral and Visorsurf.

# APPENDIX A: CALCULATION OF THE REFLECTION AND TRANSMISSION COEFFICIENTS

We consider two infinite homogeneous and isotropic chiral slabs of thickness $d$ each, as shown in Fig. A1. The first slab $(\varepsilon_B, \mu_B, \kappa_B)$ is confined between $z = -d$ and $z = 0$. The second slab $(\varepsilon_C, \mu_C, \kappa_C)$ is confined between $z = 0$ and $z = d$. A plane wave is incident at an angle $\theta_i$ upon the left chiral slab. The reflected and transmitted angles are $\theta_r$ and $\theta_t$, the refraction angles in the first chiral slab are $\theta_+^{(B)}$ and $\theta_-^{(B)}$, and in the second chiral slab are $\theta_+^{(C)}$ and $\theta_-^{(C)}$, respectively. (The subscript + applies to RCP and the – to LCP waves.)

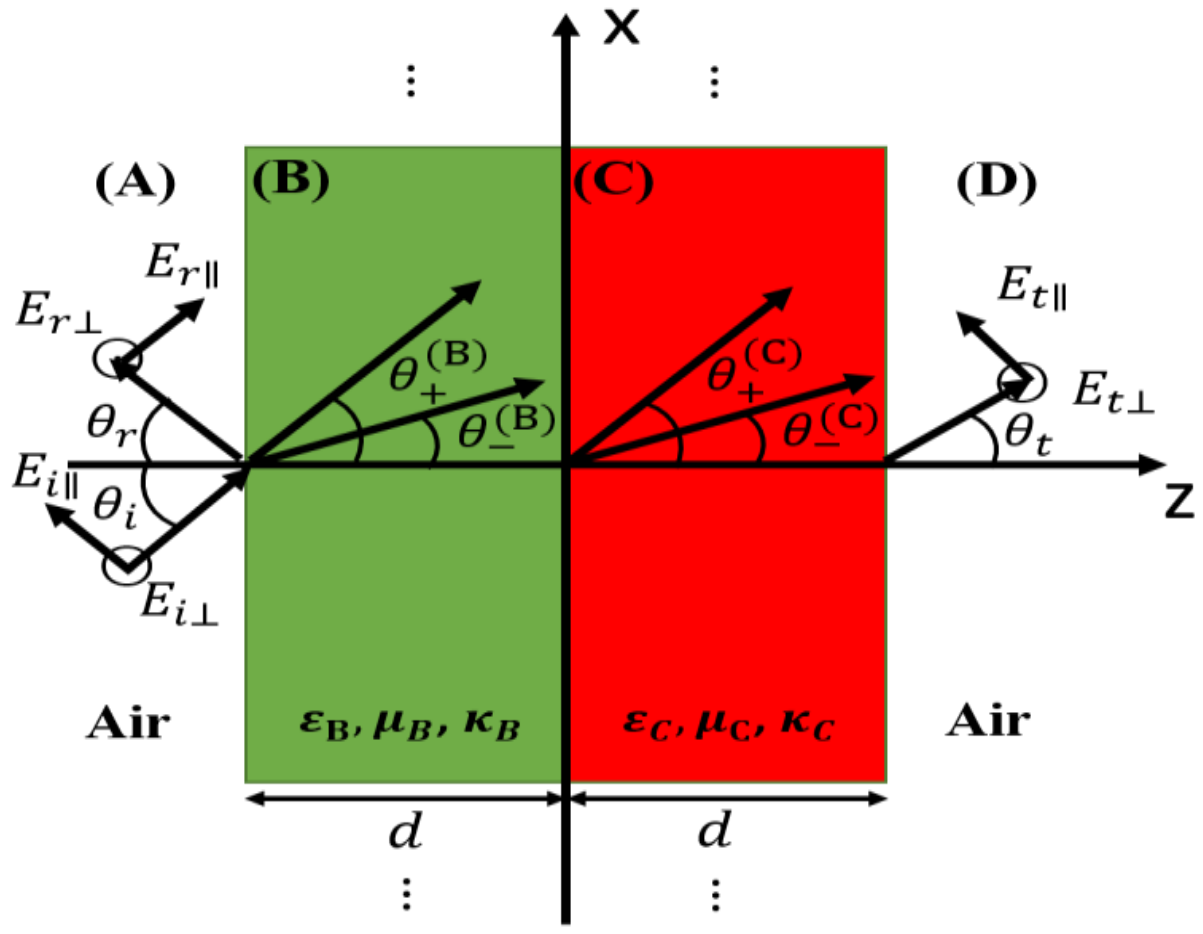

Figure A1. Problem geometry. A (1D) structure, consisting of two layers.

In the region A ($z \leq d$) the incident electric and magnetic fields can be written as (omitting the time dependence):

$$\boldsymbol{E}_i^{(A)} = [E_{i\perp}\hat{y} + E_{i\parallel}(cos\theta_i\hat{x} + sin\theta_i\hat{z})] \exp[ik_i(z\, cos\theta_i - x\, sin\theta_i)] \quad \text{(A1)}$$

$$\boldsymbol{H}_i^{(A)} = \frac{1}{\eta_A}[E_{i\parallel}\hat{y} - E_{i\perp}(cos\theta_i\hat{x} + sin\theta_i\hat{z})] \exp[ik_i(z\, cos\theta_i - x\, sin\theta_i)] \quad \text{(A2)}$$

where $\eta_A = \sqrt{\frac{\mu_A\mu_0}{\varepsilon_A\varepsilon_0}}$. Note that by setting $E_{i\perp} = 0$ we have TM waves while by setting $E_{i\parallel} = 0$ we have TE waves.The reflected electromagnetic fields can be expressed as:

$$\boldsymbol{E}_r^{(A)} = [E_{r\perp}\hat{y} + E_{r\parallel}(cos\theta_r\hat{x} - sin\theta_r\hat{z})] \exp[-ik_r(z\, cos\theta_r + x\, sin\theta_r)] \quad \text{(A3)}$$

$$\boldsymbol{H}_r^{(A)} = \frac{1}{\eta_A}[-E_{r\parallel}\hat{y} + E_{r\perp}(cos\theta_r\hat{x} - sin\theta_r\hat{z})] \exp[-ik_r(z\, cos\theta_r + x\, sin\theta_r)] \quad \text{(A4)}$$

In the region B $(-d \leq z \leq 0)$ the wave can be decomposed into four circularly polarized (CP) electromagnetic waves, two propagating towards the interface $z = 0$ and the other two propagating towards the interface $z = -d$. Thus in the first chiral slab the electromagnetic field can be represented as:

$$\boldsymbol{E}_{+}^{(\boldsymbol{B})} = E_{+}^{(B1)}\left(cos\theta_{+}^{(B)}\widehat{\boldsymbol{x}} + i\widehat{\boldsymbol{y}} + sin\theta_{+}^{(B)}\widehat{\boldsymbol{z}}\right)\exp\left[ik_{+}^{(B)}\left(z\,cos\theta_{+}^{(B)} - x\,sin\theta_{+}^{(B)}\right)\right] + E_{+}^{(B2)}\left(cos\theta_{-}^{(B)}\widehat{\boldsymbol{x}} - i\widehat{\boldsymbol{y}} + sin\theta_{-}^{(B)}\widehat{\boldsymbol{z}}\right)\exp\left[ik_{-}^{(B)}\left(z\,cos\theta_{-}^{(B)} - x\,sin\theta_{-}^{(B)}\right)\right] \quad \text{(A5)}$$

$$\boldsymbol{H}_{+}^{(\boldsymbol{B})} = \frac{-i}{Z_{-}^{(B)}}E_{+}^{(B1)}\left(cos\theta_{+}^{(B)}\widehat{\boldsymbol{x}} + i\widehat{\boldsymbol{y}} + sin\theta_{+}^{(B)}\widehat{\boldsymbol{z}}\right)\exp\left[ik_{+}^{(B)}\left(z\,cos\theta_{+}^{(B)} - x\,sin\theta_{+}^{(B)}\right)\right] + \frac{i}{Z_{-}^{(B)}}E_{+}^{(B2)}\left(cos\theta_{-}^{(B)}\widehat{\boldsymbol{x}} - i\widehat{\boldsymbol{y}} + sin\theta_{-}^{(B)}\widehat{\boldsymbol{z}}\right)\exp\left[ik_{-}^{(B)}\left(z\,cos\theta_{-}^{(B)} - x\,sin\theta_{-}^{(B)}\right)\right] \quad \text{(A6)}$$

$$\boldsymbol{E}_{-}^{(\boldsymbol{B})} = E_{-}^{(B1)}\left(-cos\theta_{+}^{(B)}\widehat{\boldsymbol{x}} + i\widehat{\boldsymbol{y}} + sin\theta_{+}^{(B)}\widehat{\boldsymbol{z}}\right)\exp\left[-ik_{+}^{(B)}\left(z\,cos\theta_{+}^{(B)} + x\,sin\theta_{+}^{(B)}\right)\right] + E_{-}^{(B2)}\left(-cos\theta_{-}^{(B)}\widehat{\boldsymbol{x}} - i\widehat{\boldsymbol{y}} + sin\theta_{-}^{(B)}\widehat{\boldsymbol{z}}\right)\exp\left[-ik_{-}^{(B)}\left(z\,cos\theta_{-}^{(B)} + x\,sin\theta_{-}^{(B)}\right)\right] \quad \text{(A7)}$$

$$\boldsymbol{H}_{-}^{(\boldsymbol{B})} = \frac{-i}{Z_{+}^{(B)}}E_{-}^{(B1)}\left(-cos\theta_{+}^{(B)}\widehat{\boldsymbol{x}} + i\widehat{\boldsymbol{y}} + sin\theta_{+}^{(B)}\widehat{\boldsymbol{z}}\right)\exp\left[-ik_{+}^{(B)}\left(z\,cos\theta_{+}^{(B)} + x\,sin\theta_{+}^{(B)}\right)\right] + \frac{i}{Z_{-}^{(B)}}E_{-}^{(B2)}\left(-cos\theta_{-}^{(B)}\widehat{\boldsymbol{x}} - i\widehat{\boldsymbol{y}} + sin\theta_{-}^{(B)}\widehat{\boldsymbol{z}}\right) \quad \text{(A8)}$$

where $Z_{\pm}^{(B)} = \sqrt{\frac{\mu_B\mu_0}{\varepsilon_B\varepsilon_0}}$ is the wave impedance in the air - material-B interface , and $k_{\pm}^{(B)} = \frac{\omega}{c}(\sqrt{\varepsilon_B\mu_B} \pm \kappa_B)$ [27,29,35] are the wavenumbers for circularly polarized waves propagating in the chiral material B.

In the region C $(0 \leq z \leq d)$ there are four CP electromagnetic waves, two propagating towards the interface $z = d$ and the other two propagating towards the interface $z = 0$. Thus in the second chiral slab the electromagnetic field can be represented as:

$$\boldsymbol{E}_{+}^{(\boldsymbol{C})} = E_{+}^{(C1)}\left(cos\theta_{+}^{(C)}\widehat{\boldsymbol{x}} + i\widehat{\boldsymbol{y}} + sin\theta_{+}^{(C)}\widehat{\boldsymbol{z}}\right)\exp\left[ik_{+}^{(C)}\left(z\,cos\theta_{+}^{(C)} - x\,sin\theta_{+}^{(C)}\right)\right] + E_{+}^{(C2)}\left(cos\theta_{-}^{(C)}\widehat{\boldsymbol{x}} - i\widehat{\boldsymbol{y}} + sin\theta_{-}^{(C)}\widehat{\boldsymbol{z}}\right)\exp\left[ik_{-}^{(C)}\left(z\,cos\theta_{-}^{(C)} - x\,sin\theta_{-}^{(C)}\right)\right] \quad \text{(A9)}$$

$$\boldsymbol{H}_{+}^{(\boldsymbol{C})} = \frac{-i}{Z_{-}^{(C)}}E_{+}^{(C1)}\left(cos\theta_{+}^{(C)}\widehat{\boldsymbol{x}} + i\widehat{\boldsymbol{y}} + sin\theta_{+}^{(C)}\widehat{\boldsymbol{z}}\right)\exp\left[ik_{+}^{(C)}\left(z\,cos\theta_{+}^{(C)} - x\,sin\theta_{+}^{(C)}\right)\right] + \frac{i}{Z_{-}^{(C)}}E_{+}^{(C2)}\left(cos\theta_{-}^{(C)}\widehat{\boldsymbol{x}} - i\widehat{\boldsymbol{y}} + sin\theta_{-}^{(C)}\widehat{\boldsymbol{z}}\right)\exp\left[ik_{-}^{(C)}\left(z\,cos\theta_{-}^{(C)} - x\,sin\theta_{-}^{(C)}\right)\right] \quad \text{(A10)}$$

$$\boldsymbol{E}_{-}^{(\boldsymbol{C})} = E_{-}^{(C1)}\left(-cos\theta_{+}^{(C)}\widehat{\boldsymbol{x}} + i\widehat{\boldsymbol{y}} + sin\theta_{+}^{(C)}\widehat{\boldsymbol{z}}\right)\exp\left[-ik_{+}^{(C)}\left(z\,cos\theta_{+}^{(C)} + x\,sin\theta_{+}^{(C)}\right)\right] + E_{-}^{(C2)}\left(-cos\theta_{-}^{(C)}\widehat{\boldsymbol{x}} - i\widehat{\boldsymbol{y}} + sin\theta_{-}^{(C)}\widehat{\boldsymbol{z}}\right)\exp\left[-ik_{-}^{(C)}\left(z\,cos\theta_{-}^{(C)} + x\,sin\theta_{-}^{(C)}\right)\right] \quad \text{(A11)}$$

$$\boldsymbol{H}_{-}^{(\boldsymbol{C})} = \frac{-i}{Z_{+}^{(C)}}E_{-}^{(C1)}\left(-cos\theta_{+}^{(C)}\widehat{\boldsymbol{x}} + i\widehat{\boldsymbol{y}} + sin\theta_{+}^{(C)}\widehat{\boldsymbol{z}}\right)\exp\left[-ik_{+}^{(C)}\left(z\,cos\theta_{+}^{(C)} + x\,sin\theta_{+}^{(C)}\right)\right] + \frac{i}{Z_{-}^{(C)}}E_{-}^{(C2)}\left(-cos\theta_{-}^{(C)}\widehat{\boldsymbol{x}} - i\widehat{\boldsymbol{y}} + sin\theta_{-}^{(C)}\widehat{\boldsymbol{z}}\right)\exp\left[-ik_{-}^{(C)}\left(z\,cos\theta_{-}^{(C)} + x\,sin\theta_{-}^{(C)}\right)\right] \quad \text{(A12)}$$

where $Z_{\pm}^{(C)} = \sqrt{\frac{\mu_C\mu_0}{\varepsilon_C\varepsilon_0}}$ , and $k_{\pm}^{(C)} = \frac{\omega}{c}\left(\sqrt{\varepsilon_C\mu_C} \pm \kappa_C\right)$.

In the region D ( $z \geq d$), the transmitted electromagnetic fields can be written as:

$$\boldsymbol{E}_{\boldsymbol{t}}^{(\boldsymbol{D})} = [E_{t\perp}\hat{y} + E_{t\parallel}(cos\theta_t\hat{x} + sin\theta_t\hat{z})]\exp[ik_t(z\,cos\theta_t - x\,sin\theta_t)] \quad \text{(A13)}$$

$$\boldsymbol{H}_{\boldsymbol{t}}^{(\boldsymbol{D})} = \frac{1}{\eta_D}[E_{t\parallel}\hat{y} - E_{t\perp}(cos\theta_t\hat{x} + sin\theta_t\hat{z})]\exp[ik_t(z\,cos\theta_t - x\,sin\theta_t)] \quad \text{(A14)}$$

Where $\eta_D = \sqrt{\frac{\mu_D\mu_0}{\varepsilon_D\varepsilon_0}}$.

According to the boundary conditions at the interfaces $z = -d, z = 0$ and $z = d$:

$$\begin{cases} \boldsymbol{E}_i^{(\boldsymbol{A})}(z=-d)\big|_T + \boldsymbol{E}_r^{(\boldsymbol{A})}(z=-d)\big|_T = \boldsymbol{E}_{+}^{(B)}(z=-d)\big|_T + \boldsymbol{E}_{-}^{(B)}(z=-d)\big|_T \\ \boldsymbol{H}_i^{(\boldsymbol{A})}(z=-d)\big|_T + \boldsymbol{H}_r^{(\boldsymbol{A})}(z=-d)\big|_T = \boldsymbol{H}_{+}^{(B)}(z=-d)\big|_T + \boldsymbol{H}_{-}^{(B)}(z=-d)\big|_T \\ \boldsymbol{E}_{+}^{(B)}(z=0)\big|_T + \boldsymbol{E}_{-}^{(B)}(z=0)\big|_T = \boldsymbol{E}_{+}^{(C)}(z=0)\big|_T + \boldsymbol{E}_{-}^{(C)}(z=0)\big|_T \\ \boldsymbol{H}_{+}^{(B)}(z=0)\big|_T + \boldsymbol{H}_{-}^{(B)}(z=0)\big|_T = \boldsymbol{H}_{+}^{(C)}(z=0)\big|_T + \boldsymbol{H}_{-}^{(C)}(z=0)\big|_T \\ \boldsymbol{E}_{+}^{(C)}(z=d)\big|_T + \boldsymbol{E}_{-}^{(C)}(z=d)\big|_T = \boldsymbol{E}_t^{(\boldsymbol{D})}(z=d)\big|_T \\ \boldsymbol{H}_{+}^{(C)}(z=d)\big|_T + \boldsymbol{H}_{-}^{(C)}(z=d)\big|_T = \boldsymbol{H}_t^{(\boldsymbol{D})}(z=d)\big|_T \end{cases} . \quad \text{(A15)}$$

where $|_T$ represents tangential components of the electromagnetic fields.

Equating the tangential components of $\boldsymbol{E}$ and $\boldsymbol{H}$ $(E_x, E_y, H_x, H_y)$ at the three structure interfaces results to a 12x12 system of linear equations. For convenience, we write it in a matrix form, as

$$
\begin{pmatrix}
0 & -e^{i\delta_{Bi}} & R_{B1}e^{-i\delta_{B1}} & R_{B2}e^{-i\delta_{B2}} & -R_{B1}e^{i\delta_{B1}} & -R_{B2}e^{i\delta_{B2}} & 0 & 0 & 0 & 0 & 0 & 0 \\
-e^{i\delta_{Bi}} & 0 & ie^{-i\delta_{B1}} & -ie^{-i\delta_{B2}} & ie^{i\delta_{B1}} & -ie^{i\delta_{B2}} & 0 & 0 & 0 & 0 & 0 & 0 \\
e^{i\delta_{Bi}} & 0 & i\frac{\eta_A}{Z_B}R_{B1}e^{-i\delta_{B1}} & -i\frac{\eta_A}{Z_B}R_{B2}e^{-i\delta_{B2}} & -i\frac{\eta_A}{Z_B}R_{B1}e^{i\delta_{B1}} & i\frac{\eta_A}{Z_B}R_{B2}e^{i\delta_{B2}} & 0 & 0 & 0 & 0 & 0 & 0 \\
0 & e^{i\delta_{Bi}} & \frac{\eta_A}{Z_B}e^{-i\delta_{B1}} & \frac{\eta_A}{Z_B}e^{-i\delta_{B2}} & \frac{\eta_A}{Z_B}e^{i\delta_{B1}} & \frac{\eta_A}{Z_B}e^{i\delta_{B2}} & 0 & 0 & 0 & 0 & 0 & 0 \\
0 & 0 & \cos\theta_+^{(B)} & \cos\theta_-^{(B)} & -\cos\theta_+^{(B)} & -\cos\theta_-^{(B)} & -\cos\theta_+^{(C)} & -\cos\theta_-^{(C)} & cos\theta_+^{(C)} & cos\theta_-^{(C)} & 0 & 0 \\
0 & 0 & i & -i & i & -i & -i & i & -i & i & 0 & 0 \\
0 & 0 & \frac{-i\cos\theta_+^{(B)}}{Z_B} & \frac{i\cos\theta_-^{(B)}}{Z_B} & \frac{i\cos\theta_+^{(B)}}{Z_B} & \frac{-i\cos\theta_-^{(B)}}{Z_B} & \frac{i\cos\theta_+^{(C)}}{Z_C} & \frac{-i\cos\theta_-^{(C)}}{Z_C} & \frac{-i\cos\theta_+^{(C)}}{Z_C} & \frac{i\cos\theta_-^{(C)}}{Z_C} & 0 & 0 \\
0 & 0 & \frac{1}{Z_B} & \frac{1}{Z_B} & \frac{1}{Z_B} & \frac{1}{Z_B} & -\frac{1}{Z_C} & -\frac{1}{Z_C} & -\frac{1}{Z_C} & -\frac{1}{Z_C} & 0 & 0 \\
0 & 0 & 0 & 0 & 0 & 0 & R_{C1}e^{i\delta_{C1}} & R_{C2}e^{i\delta_{C2}} & -R_{C1}e^{-i\delta_{C1}} & -R_{C2}e^{-i\delta_{C2}} & 0 & -e^{i\delta_{Ct}} \\
0 & 0 & 0 & 0 & 0 & 0 & ie^{i\delta_{C1}} & -ie^{i\delta_{C2}} & ie^{-i\delta_{C1}} & -ie^{-i\delta_{C2}} & -e^{i\delta_{Ct}} & 0 \\
0 & 0 & 0 & 0 & 0 & 0 & -i\frac{\eta_D}{Z_C}R_{C1}e^{i\delta_{C1}} & i\frac{\eta_D}{Z_C}R_{C2}e^{i\delta_{C2}} & i\frac{\eta_D}{Z_C}R_{C1}e^{-i\delta_{C1}} & -i\frac{\eta_D}{Z_C}R_{C2}e^{-i\delta_{C2}} & e^{i\delta_{Ct}} & 0 \\
0 & 0 & 0 & 0 & 0 & 0 & \frac{\eta_D}{Z_C}e^{i\delta_{C1}} & \frac{\eta_D}{Z_C}e^{i\delta_{C2}} & \frac{\eta_D}{Z_C}e^{-i\delta_{C1}} & \frac{\eta_D}{Z_C}e^{-i\delta_{C2}} & 0 & -e^{i\delta_{Ct}}
\end{pmatrix}
\begin{pmatrix}
E_{r\perp} \\ E_{r\parallel} \\ E_+^{(B1)} \\ E_+^{(B2)} \\ E_-^{(B1)} \\ E_-^{(B2)} \\ E_+^{(C1)} \\ E_+^{(C2)} \\ E_-^{(C1)} \\ E_-^{(C2)} \\ E_{t\perp} \\ E_{t\parallel}
\end{pmatrix}
=
\begin{pmatrix}
E_{i\parallel}\, e^{-i\delta_{Bi}} \\ E_{i\perp}e^{-i\delta_{Bi}} \\ E_{i\perp}e^{-i\delta_{Bi}} \\ E_{i\parallel}e^{-i\delta_{Bi}} \\ 0 \\ 0 \\ 0 \\ 0 \\ 0 \\ 0 \\ 0 \\ 0
\end{pmatrix}
\qquad \text{(A16)}
$$

Where $\delta_{Bi} = k_i dcos\theta_i, \delta_{B1} = k_+^{(B)} dcos\theta_+^{(B)}, \delta_{B2} = k_-^{(B)} dcos\theta_-^{(B)}, \delta_{C1} = k_+^{(C)} dcos\theta_+^{(C)},\ \ \delta_{C2} = k_-^{(C)} dcos\theta_-^{(C)}, \delta_{Ct} = k_t dcos\theta_t$
$R_{B1} = \frac{cos\theta_+^{(B)}}{cos\theta_i}, R_{B2} = \frac{cos\theta_-^{(B)}}{cos\theta_i}, R_{C1} = \frac{cos\theta_+^{(C)}}{cos\theta_i}, R_{C2} = \frac{cos\theta_-^{(C)}}{cos\theta_i},$
and $k_i = \omega\sqrt{\varepsilon_A \mu_A}\ ,\ \ k_t = \omega\sqrt{\varepsilon_D \mu_D}.$

Since the analytical solution of this system of twelve equations leads to complicated expressions for the field amplitudes, we resort to numerical techniques to invert the matrix equation (A16). Note that by setting $E_{i\perp} = 0$ in Eq. (A16) we have TM waves while by setting $E_{i\parallel} = 0$ we have TE waves, respectively.
From the solution of the above 12x12 system one can obtain the reflection and transmission coefficients for TM and TE polarized incident waves as:

$$
\begin{pmatrix} E_{t\perp} \\ E_{t\parallel} \end{pmatrix} = \begin{pmatrix} t_{\perp\perp} & t_{\perp\parallel} \\ t_{\parallel\perp} & t_{\parallel\parallel} \end{pmatrix} \begin{pmatrix} E_{i\perp} \\ E_{i\parallel} \end{pmatrix} = T_{lin} \begin{pmatrix} E_{i\perp} \\ E_{i\parallel} \end{pmatrix}, \qquad \text{(A17)}
$$

and

$$
\begin{pmatrix} E_{r\perp} \\ E_{r\parallel} \end{pmatrix} = \begin{pmatrix} r_{\perp\perp} & r_{\perp\parallel} \\ r_{\parallel\perp} & r_{\parallel\parallel} \end{pmatrix} \begin{pmatrix} E_{i\perp} \\ E_{i\parallel} \end{pmatrix} = R_{lin} \begin{pmatrix} E_{i\perp} \\ E_{i\parallel} \end{pmatrix}. \qquad \text{(A18)}
$$

For analysing the transmission and reflection properties of a *PT*-chiral system is advantageous to also have at hand the transmission, *T*, and reflection, *R*, matrices for circularly polarized waves. They can be obtained from Eqs. (A17), (A18) by a change of the base vectors [27,29], as

$$
T_{cir} = \frac{1}{2}\begin{pmatrix} (t_{\parallel\parallel} + t_{\perp\perp}) + i(t_{\parallel\perp} - t_{\perp\parallel}) & (t_{\parallel\parallel} - t_{\perp\perp}) - i(t_{\parallel\perp} + t_{\perp\parallel}) \\ (t_{\parallel\parallel} - t_{\perp\perp}) + i(t_{\parallel\perp} + t_{\perp\parallel}) & (t_{\parallel\parallel} + t_{\perp\perp}) - i(t_{\parallel\perp} - t_{\perp\parallel}) \end{pmatrix} \qquad \text{(A19)}
$$

and

$$
R_{cir} = \frac{1}{2}\begin{pmatrix} (r_{\parallel\parallel} - r_{\perp\perp}) + i(r_{\parallel\perp} + r_{\perp\parallel}) & (r_{\parallel\parallel} + r_{\perp\perp}) - i(r_{\parallel\perp} - r_{\perp\parallel}) \\ (r_{\parallel\parallel} + r_{\perp\perp}) + i(r_{\parallel\perp} - r_{\perp\parallel}) & (r_{\parallel\parallel} - r_{\perp\perp}) - i(r_{x\perp} + r_{\perp\parallel}) \end{pmatrix} \qquad \text{(A20)}
$$

$T_{cir}$ and $R_{cir}$ connect the amplitudes of circularly polarized incident waves and scattered waves:

$$\begin{pmatrix} E_{t+} \\ E_{t-} \end{pmatrix} = T_{cir} \begin{pmatrix} E_{i+} \\ E_{i-} \end{pmatrix} = \begin{pmatrix} t_{++} & t_{+-} \\ t_{-+} & t_{--} \end{pmatrix} \begin{pmatrix} E_{i+} \\ E_{i-} \end{pmatrix}, \tag{A21}$$

and

$$\begin{pmatrix} E_{r+} \\ E_{r-} \end{pmatrix} = R_{cir} \begin{pmatrix} E_{i+} \\ E_{i-} \end{pmatrix} = \begin{pmatrix} r_{++} & r_{+-} \\ r_{-+} & r_{--} \end{pmatrix} \begin{pmatrix} E_{i+} \\ E_{i-} \end{pmatrix}. \tag{A22}$$

The reflection and transmission coefficients when the incident wave is from the right side of the slab shown in Fig. A1 can be obtained from above analysis by exchanging the material parameters in the two layers.

## APPENDIX B: POLARIZATION ANALYSIS – STOKES PARAMETERS

Taking into account the Stokes parameters, we calculated the ellipticity angle ($\chi$) and the orientation angle of the polarization ellipse ($\psi$) (see Fig. B1) of the wave transmitted through the chiral structure. This approach is applied to any electromagnetic wave of the form

$$\boldsymbol{E} = [E_{\perp}\hat{y} + E_{\parallel}(cos\theta\hat{x} + sin\theta\hat{z})] \exp[i(q - \omega t)] \tag{B1}$$

where $q = k(z\,cos\theta - x\,sin\theta)$ and $\hat{x}, \hat{y}, \hat{z}$ the unit vectors. For the wave of Eq. (B1) the four Stokes parameters, which describe completely the state of polarization [34], are given by

$$S_0 = E_{\perp}E_{\perp}^* + E_{\parallel}E_{\parallel}^* \tag{B2a}$$
$$S_1 = E_{\perp}E_{\perp}^* - E_{\parallel}E_{\parallel}^* \tag{B2b}$$
$$S_2 = 2Re[E_{\perp}E_{\parallel}^*] \tag{B2c}$$
$$S_3 = 2Im[E_{\perp}E_{\parallel}^*] \tag{B2d}$$

with $^*$ denoting the complex conjugate. Through $S$ parameters the ellipticity angle is given by

$$\tan\chi = \frac{S_3/S_0}{1+[1-(S_3/S_0)^2]^{1/2}},\ (-\pi/4 \leq \chi \leq \pi/4) \tag{B3}$$

and the orientation angle of the polarization ellipse ($\psi$) (see Fig. B1) is given by

$$\tan 2\psi = \frac{S_2}{S_1},\ (0 \leq \psi \leq \pi) \tag{B4}$$

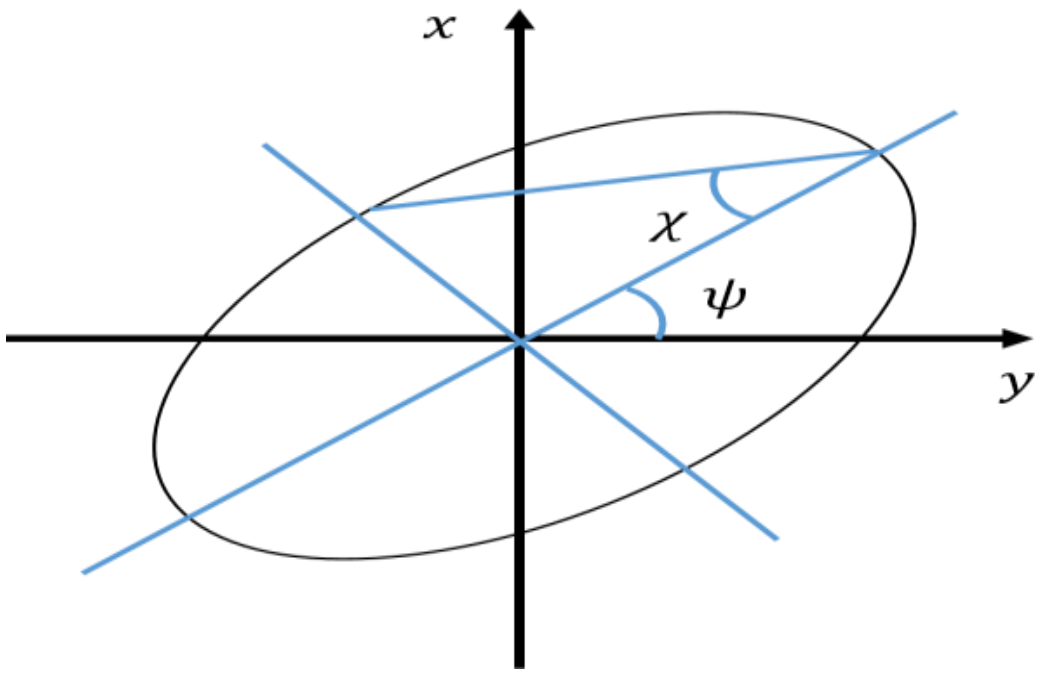

Figure B1. The polarization ellipse for an elliptically polarized plane wave with an ellipticity ($\chi$) and an orientation angle ($\psi$).

## APPENDIX C: *PT*-SYMMETRIC PHASES VERSUS CHIRALITY

In order to characterize further the impact of chirality on the *PT*-symmetric phases of the chiral *PT*-symmetric system investigated in this work, we calculate the eigenvalues of scattering matrix as a function of: (a) real part of chirality (see Fig. C1 (a)) keeping constant the imaginary part and (b) imaginary part of chirality (see Fig. C1 (b)) keeping constant the real part. In Fig. C1, we show that it is possible to control the *PT*-phases either with only the real or with only the imaginary part of the chirality. This possibility gives us an additional degree of freedom in controlling the different *PT*-related phases and the associated exceptional points. Here, we assume the same material parameters as in a manuscript.

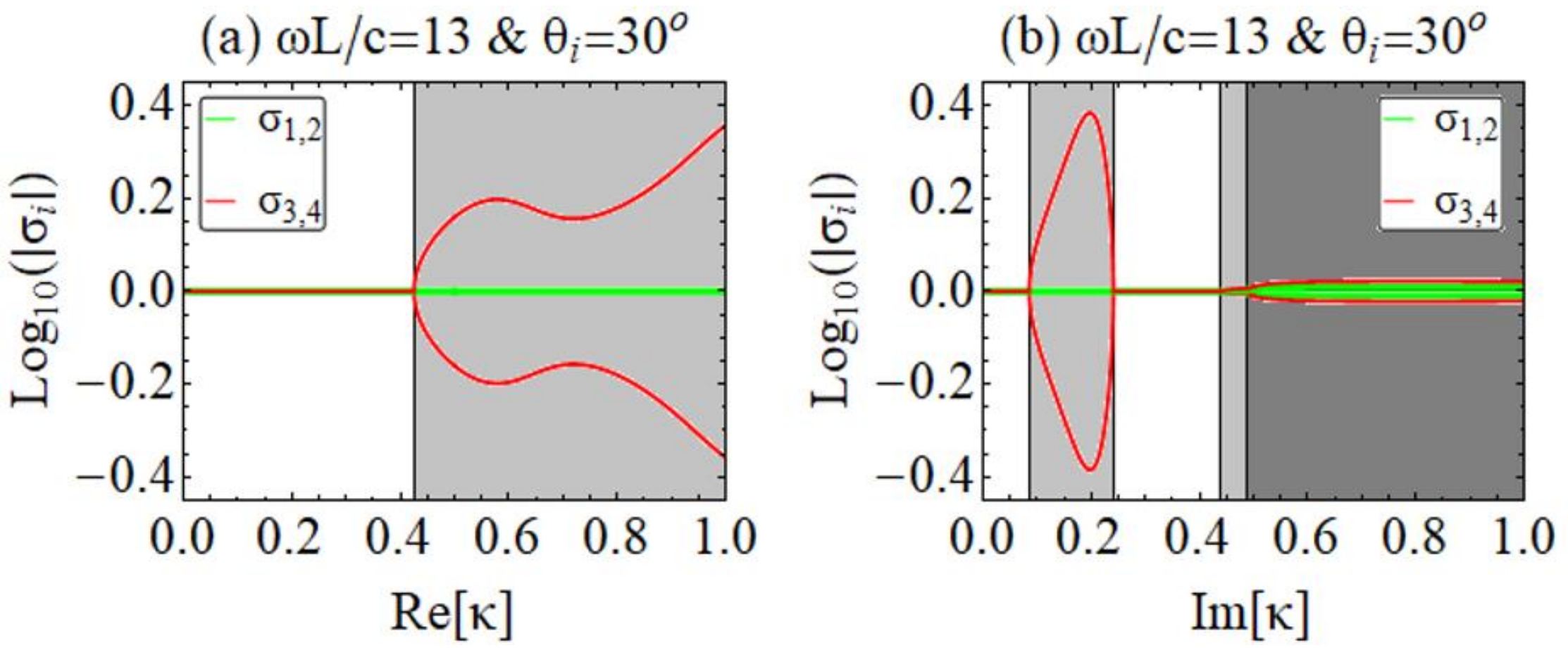

Figure C1. The eigenvalues of $S$ matrix as a function of: (a) real part of the chirality parameter ($\kappa(-z) = -Re[\kappa] + 0.05i$, $\kappa(z) = Re[\kappa] + 0.05i$) and (b) imaginary part of the chirality ($\kappa(-z) = -0.05 + Im[\kappa]i$, $\kappa(z) = 0.05 + Im[\kappa]i$), for obliquely incident waves at $\theta_i = 30^o$ and for dimensionless frequency $\frac{\omega L}{c} = 13.0$. The material parameter of the two media are: $\varepsilon(-z) = 4 - 0.6i$, $\varepsilon(z) = 4 + 0.6i$, $\mu(-z) = 1.5 - 0.15i$, $\mu(z) = 1.5 + 0.15i$.

* katsantonis@iesl.forth.gr
† sdroulias@iesl.forth.gr
‡ kafesaki@iesl.forth.gr

**References**

[1] C. M. Bender and S. Boettcher, Phys. Rev. Lett. **80**, 5243 (1998).
[2] C.M. Bender, S. Boettcher, and P.N. Meisinger, J. Math. Phys. **40**, 2201 (1999).
[3] C. M. Bender, D. C. Brody, and H. F. Jones, Phys. Rev. Lett. **89**, 270401 (2002).
[4] C. M. Bender, Rep. Prog. Phys. **70**, 947 (2007).
[5] R. El-Ganainy, K. G. Makris, D. N. Christodoulides, and Z. H. Musslimani, Opt. Lett. **32**, 2632-2634 (2007).
[6] K. G. Makris, R. El-Ganainy, D. N. Christodoulides, and Z. H. Musslimani, Phys. Rev. Lett. **100**, 103904 (2008).
[7] A. Guo, G. J. Salamo, D. Duchesne, R.Morandotti, M. Volatier-Ravat, V. Aimez, G. A. Siviloglou, and D. N. Christodoulides, Phys. rev. Lett. **103**, 093902 (2009).
[8] E. Rüter, K.G. Makris, R.El-Ganainy, D. N. Christodoulides, M. Sagev, and D. Kip, Nat. Phys. **6**, 192 (2010).
[9] L. Feng, M. Ayache, J. Huang, Y. Xu, M. Lu, Y. Chen, Y. Fainman, and A. Scherer, Science **333**, 729 (2011).
[10] L. Feng, Y. L. Xu, W. S. Fegadolli, M. H. Lu, J. E. Oliveira,V. R. Almeida, Y. F. Chen, and A. Scherer, Nat. Mater. **12**, 108 (2013).
[11] Y. Shen, X. Hua Deng, and L. Chen, Opt. Express **22**, 19440 (2014).
[12] V. Yannopapas, Phys. Rev. A **89**, 013808 (2014).
[13] Li Ge, Y.D. Chong and A. D. Stone, Phys. Rev. A **85**, 023802 (2012).
[14] Z. Lin, H. Ramezani, T. Eichelkraut, T. Kottos, H. Cao, and D. N. Cristodoulides, Phys. Rev. Lett. **106**, 213901 (2011).
[15] Y. D. Chong, L. Ge, and A. D. Stone, Phys. Rev. Lett. **106**, 093902 (2011).
[16] Y. Sun, W. Tan, H.Q. Li, J. Li, and H. Chen, Phys. Rev. Lett. **112**, 143903 (2014).
[17] N. Lazarides and G. P. Tsironis, Phys. Rev. Lett. **110**, 053901 (2013).
[18] G. Castaldi, S. Savoia, V. Galdi, A. Alu and N. Engheta, Phys. Rev. Lett. **110**, 173901 (2013).
[19] S. Savoia, G. Castaldi, V. Galdi, A. Alu and N. Engheta, Phys. Rev. B **89**, 085105 (2014).
[20] S. Savoia, G. Castaldi, V. Galdi, A. Alu and N. Engheta, Phys. Rev. B **91**, 115114 (2015).
[21] J. Wang, H. Dong, C. Ling, C. T. Chan, and K. H. Fung, Phys. Rev. B **91**, 235410 (2015).
[22] O. V. Shramkova and G. P. Tsironis, J. Opt. **18**, 105101, (2016).
[23]Y. Y. Fu, Y. D. Xu, and H. Y. Chen, Opt. Exp., **24** (2), 1648–1657 (2016).
[24] A. Krasnok, D. Baranov, H. Li, M. Ali Miri, F. Monticone, A. Alu, Adv. Opt. Photonics, **11**, 892 (2019).
[25] S. Droulias, I. Katsantonis, M. Kafesaki, C. M. Soukoulis and E.N. Economou, Phys. Rev. B **100**, 205133 (2019).
[26] S. Droulias, I. Katsantonis, M. Kafesaki, C. M. Soukoulis and E.N. Economou, Phys. Rev. Lett. **122**, 213201 (2019).
[27] I. V. Lindell, A. H. Sihvola, S. A. Tretyakov and A. J. Viitanen, (Boston, MA: Artech House Publishers) (1994).
[28] Applequist J. Optical activity: Biot's bequest Am. Sci. **75** 58–68 (1987).
[29] G. Kenanakis, R. Zhao, N. Katsarakis, M. Kafesaki, C. M. Soukoulis, and E. N. Economou, Opt. Express **22**, 12149 (2014).
[30] S. Tretyakov, I. Nefedov, A. Sihvola, S. Maslovski, and C. Simovski, J. Electromagn. Waves Appl. **17**, 695 (2003).
[31] J. B. Pendry, Science **306**, 1353 (2004).
[32] E. Plum, J. Zhou, J. Dong, V. A. Fedotov, T. Koschny, C. M. Soukoulis and N. I. Zheludev, Phys. Rev. B **79**, 035407 (2009).
[33] J. Zhou, J. Dong, B.Wang, T. Koschny, M. Kafesaki and C. M. Soukoulis, Phys. Rev. B **79**, 121104 (2009).
[34] Bassiri, S.; Papas, C. H.; Engheta, N., J. Opt. Soc. Am. A 1988, 5, 1450-1459.
[35] B. Wang, J. Zhou, T. Koschny, M. Kafesaki and C. M. Soukoulis, J. Opt. A **11**, 114003 (2009).
[36] R. Shankar, Principles of Quantum Mechanics (Plenum Press, 1994).
[37] C. Altman and K. Suchy, Reciprocity, Spatial Mapping and Time Reversal in Electromagnetics (Springer, 2014)
[38] Caloz, C.; Alù, A.; Tretyakov, S.; Sounas, D.; Achouri, K.; Deck-Léger, Z.L. Phys. Rev. Appl. , **10**, 047001 (2018).
[39] A. A. Zyablovsky, A. P. Vinogradov, A. A. Pukhov, A. V. Dorofeenko, and A. A. Lisyansky, Phys. Usp. **57**,1063 2014.
[40] W. Chen, Ş. K. Özdemir, G. Zhao, J. Wiersig, and L. Yang, Nature **548**, 192 (2017).
[41] Q. Zhong, J. Ren, M. Khajavikhan, D. N. Christodoulides, Ş. K. Özdemir, and R. El-Ganainy, Phys. Rev. Lett. , **122**, 153902 (2019).